\documentclass[aps, prd,preprint,showpacs,superscriptaddress,groupedaddress,nofootinbib]{revtex4-1}
\usepackage{amsmath}
\usepackage[dvips]{graphicx}
\usepackage{morefloats}
\graphicspath{{Fig/}}
\begin{document}

\title{Dynamics of tilted Bianchi models \\ of types~III,~IV,~V in presence of diffusion}
\author{Dmitry Shogin}
\email{dmitry.shogin@uis.no}
\affiliation{Faculty of Science and Technology, University of Stavanger, N-4036 Stavanger, Norway}
\author{Sigbj\o rn Hervik}
\email{sigbjorn.hervik@uis.no}
\affiliation{Faculty of Science and Technology, University of Stavanger, N-4036 Stavanger, Norway}

\begin{abstract}
We investigate the three types of class~B Bianchi cosmologies filled with a tilted perfect fluid undergoing velocity diffusion in a scalar field background. We consider the two most important cases: dust and radiation. A complete numerical integration of the Einstein field equations coupled with the diffusion equations is done to demonstrate how the presence of 
diffusion can affect the dynamics of cosmological evolution, where the most attention is paid to changes to the late-time behaviour. We show that aside from quantitative effects, diffusion can result in significant qualitative differences. For example, the cosmologies may recollapse if diffusion is sufficiently strong, or evolve towards the de Sitter state otherwise. In constrast to the diffusionless case, radiation isotropizes in presence of diffusion, and the tilt decreases exponentially at later times:~${V\sim e^{-0.25\tau}}$; also, we determine the decay rates of energy density, which become slower when the diffusion term is non-zero. 
\end{abstract}

\pacs{98.80.Jk, 04.40.Nr}

\maketitle

\section{Introduction}
\label{Sec:Intro}

In recent years a lot of efforts have been done to investigate diffusional effects both in special and general relativity (see e.\,g.~\cite{Herrmann2009,Herrmann2010,Haba2009,Risken1996,Calogero2011,Calogero2012,Calogero2013,Shogin2013,Shogin2014a} and references therein). In cosmology, diffusion is not considered to be a fundamental phenomenon, but rather a model to describe the interaction between two different media. In one of the simplest situations these media can be represented by a perfect fluid and a cosmological scalar field \cite{Calogero2011,Calogero2012,Calogero2013,Shogin2013,Shogin2014a}. A common assumption for cosmology is that the role of fluid particles is played by galaxies in space.
\par 
It was shown in \cite{Calogero2011} how the concept of cosmological constant can be generalized to take diffusion into account. In this case, the cosmological fluid is undergoing velocity diffusion on a scalar field background, which may be associated with cosmic dark energy~\cite{Calogero2013}. The authors investigated a simple FRW~cosmology and demostrated that presence of diffusion can lead to some interesting effects which do not appear in the standard diffusionless models with a cosmological constant.
\par 
In \cite{Shogin2013} we followed \cite{Calogero2011} and considered a plane symmetric~$G_2$ cosmology with diffusion to get a hint on diffusional effects in spatially inhomogeneous and anisotropic cosmologies. We used the orthonormal frame approach \cite{Elst1997} and derived the diffusional corrections to the equations governing the dynamics of the universe. The solutions of Einstein field equations coupled with the diffusion equations were studied numerically, and the discovered effects were described. 
\par 
A recent paper~\cite{Shogin2014a} is devoted to investigation of a general Bianchi type~VIII model with diffusion with an emphasis on analytical methods. In particular, we described some essential features of Bianchi cosmologies with diffusion, found the future attractors and determined the late-time asymptotics of such universes. The current paper is a logical continuation of \cite{Shogin2013} and can be considered as a satellite paper to~\cite{Shogin2014a}. In contrast to the latter, in the given work we concentrate our attention on numerical methods of investigation to show how diffusion can affect the evolution process of Bianchi models.
\par 
The systematic study of anisotropic Bianchi cosmologies started with a pioneering work of Ellis and MacCallum \cite{Ellis1969}, where the authors applied the general orthonormal frame method to cosmology. Since then, a lot of investigations, both analytical and numerical, have been done. In particular, attention has been paid to cosmologies of types III=VI$_{-1}$~\cite{Hervik2007,Coley2008},~IV~\cite{Hervik2005}, and~V~\cite{Collins1979,Hewitt1992,Harnett1996}, belonging to class~B. It has been shown that in the general case without a cosmological constant these three cosmologies demonstrate completely different behaviour. Also, allowing for a tilted fluid can result in some new interesting phenomena. Considering Bianchi models with tilted fluid taking part in diffusional interactions may be the next step to a deeper understanding of the dynamical properties of anisotropic universes. Bianchi type~V universes, containing the open FRW model~\cite{Wainwright1997}, are of special interest.
\par
The paper is organised as follows. In section~\ref{Sec:DiffusionModel} we briefly describe the diffusion model and explain how the diffusion is coupled to the Einstein field equations. The dimensionless scale-independent variables are introduced in section~\ref{Sec:DimensionlessVars}. In section~\ref{Sec:EqsCons} we write down the equations and constraints governing the dynamics of the considered cosmological models and also discuss the initial conditions. Section~\ref{Sec:Results} is devoted to results of analytical and numerical investigations. Summary is presented in section~\ref{Sec:Summary}.

\section{The diffusion model}
\label{Sec:DiffusionModel}

Here we shall outline the main idea behind the model of diffusion proposed in \cite{Calogero2011,Calogero2012,Calogero2013}. The geometry of the spacetime is described by the Einstein field equations
\begin{equation}
R_{\alpha\beta}-\frac{1}{2}Rg_{\alpha\beta}=\mathcal{T}_{\alpha \beta},
\end{equation}
assuming~$8\pi G=c=1.$ To satisfy the condition~$\nabla_\alpha \mathcal{T}^{\alpha \beta}=0$, imposed by the Bianchi identities, we consider a two-component energy-momentum tensor
\begin{equation}
\mathcal{T}_{\alpha \beta} = T_{\alpha \beta} + \tilde{T}_{\alpha \beta}, \label{Eq:Model:CompositeTensor}
\end{equation}
where the first part~$T_{\alpha \beta}$ is the energy-momentum tensor of the perfect fluid, while the second one~$\tilde{T}_{\alpha \beta}=-\phi g_{\alpha \beta}$ corresponds to the scalar field. The interaction between the two systems is governed by the diffusion equations: 
\begin{align}
\nabla_\alpha T^{\alpha \beta} & =\nabla_\alpha (\phi g^{\alpha \beta})=DJ^\beta, \\
J^\alpha &= k\hat{u}^\alpha,\\
\nabla_\alpha J^\alpha &= 0. \label{Eq:Model:Conservation}
\end{align}
Here~$D>0$ is the dimensionful constant of diffusion,~$J^\alpha$ is the current density,~$k$ is the number of fluid particles per unit volume, and~$\hat{u}^\alpha$ is the 4-velocity of the fluid defined by the fluid flow.
\par 
The scalar potential~$\phi$ representing vacuum energy is in fact a generalization of the cosmological constant. Namely,~$\phi$, if assumed to be positive, reduces to "ordinary" cosmological constant in the absence of diffusion~($D=0$) or in vacuum~($T_{\alpha \beta}=0$). It is therefore logical to write the modified field equations with the~$\phi$-term on the left-hand side:
\begin{equation}
R_{\alpha \beta}-\frac{1}{2}R g_{\alpha \beta}+\phi g_{\alpha \beta}=T_{\alpha \beta}.
\end{equation}
The energy-momentum tensor of the perfect fluid~$T_{\alpha \beta}$ is given by
\begin{equation}
T_{\alpha \beta}=(\hat{p}+\hat{\rho})\hat{u}_\alpha \hat{u}_\beta+\hat{p}g_{\alpha\beta},
\end{equation}
where~$\hat{p}$ and~$\hat{\rho}$ stand respectively for pressure and density of the fluid and are measured by the observers following the fluid flow. We choose the fluid to obey the barotropic equation of state:
\begin{equation}
\hat{p}=(\gamma-1)\hat{\rho},
\end{equation}
where $0<\gamma<2$ is a constant defining the kind of the fluid. The two most important cases are those of dust~($\gamma=1$) and radiation~($\gamma=4/3$).
\par 
In this paper we are considering cosmologies where the 4-velocity of the fluid is not parallel with the hypersurface-orthogonal Gaussian normal vector~$\mathbf{n}=\partial/\partial \mathbf{t}$~\cite{King1973}; $t$ is by this defined as the proper time of observers whose worldlines are orthogonal to the surfaces of homogeneity. We have chosen the fundamental observers to follow the congruences defined by the vector field~$n^\alpha$ to avoid singular behaviour which is possible for observers comoving with the fluid flow \cite{Collins1979}.

\section{Dimensionless variables}
\label{Sec:DimensionlessVars}

Within the orthonormal frame approach~\cite{Ellis1969,Elst1997}, the geometry of spatially homogeneous cosmologies, in particular the Bianchi models~\cite{Wainwright1997,Groen2007,Ellis2012}, is described in terms of the following dimensionful variables: shear variables~$\sigma_{ab}$; curvature variables~$a_c$ and~$n_{ab}$; and angular velocity~$\Omega^a$. Furthermore, we use the spatial frame with~$a_c=(a,0,0)$; and the angular velocity~$\Omega^a$ is chosen to satisfy 
\begin{equation}
\Omega^2=-\sigma_{13}, ~~ \Omega^3=\sigma_{12}. 
\end{equation}
The component~$\Omega^1$ is left undetermined at this stage and is used later to eliminate the remaining gauge freedom~\cite{Coley2005}.
\par 
In the dynamical systems method~\cite{Wainwright1997} it is common to introduce dimensionless, scale-independent variables. We choose the Hubble scalar~$H$ to be the normalization factor and introduce the normalized geometrical variables by
\begin{equation}
(\Sigma_{ab}, A, N_{ab}, R^1)=(\sigma_{ab}, a, n_{ab}, \Omega^1)/H.
\end{equation} 
The Hubble-normalized shear~$\Sigma_{ab}$ and~curvature~$N_{ab}$ are then parametrized as follows:
\begin{align}
\Sigma_{ab} &=\left[ 
\begin{array}{ccc}
-2\Sigma_+ & \sqrt{3}\Sigma_{12} & \sqrt{3}\Sigma_{13} \\
\sqrt{3}\Sigma_{12} & \Sigma_+ +\sqrt{3}\Sigma_- & \sqrt{3}\Sigma_{23} \\
\sqrt{3}\Sigma_{13} & \sqrt{3}\Sigma_{23} & \Sigma_+-\sqrt{3}\Sigma_-
\end{array} \right],\\
N_{ab} &=\sqrt{3}\left[ 
\begin{array}{ccc}
0 & 0 & 0 \\
0 & \bar{N}+N_- & N_{23} \\
0 & N_{23} & \bar{N}-N_-
\end{array} \right].
\end{align} 
We write the normalized variables as functions of dimensionless time~$\tau$. The latter is related to the proper time~$t$ by
\begin{equation}
\frac{\text{d}\tau}{\text{d}t}=H,
\end{equation}
It is also convenient to introduce the deceleration parameter~$q$ by
\begin{equation}
q=-\frac{1}{H^2}\frac{\text{d}H}{\text{d}t}-1,
\end{equation}
so that 
\begin{equation}
H^\prime = -(1+q)H,
\end{equation}
where the prime denotes the derivative over~$\tau$.
\par 

The variables describing the energy-matter content of the models are: energy density of the fluid~$\rho$; scalar potential~$\phi$; fluid 3-velocity~$v_a$; and number density of particles~$k$. The variables~$v_a$ are often called the tilt variables. If we introduce the magnitude of the tilt by~$V^2=v_av^a,$ then in case~$0<V<1$ the model is tilted; special cases~$V=0$ and~$V=1$ are called non-tilted and extremely tilted, respectively.  
\par 
The fluid variables are normalized by 
\begin{equation}
(\Omega,\Phi)=(\rho,\phi)/3H^2, ~~~ K=Dk/3H^3. \label{Eq:DLVars:FluidVars}
\end{equation}
Note that by~(\ref{Eq:DLVars:FluidVars}) the number density of particles~$k$ and the constant of diffusion~$D$ are now encapsulated into a single variable, namely the diffusion term~$K$.

\section{Evolution equations and constraints}
\label{Sec:EqsCons}

\subsection{Equations}
\label{Sec:EqsCons:Eqs}
Following \cite{Hervik2006} and~\cite{Shogin2014a}, we introduce the following complex variables:
\begin{align}
\begin{split}
\mathbf{N}_\times &= N_-+iN_{23}; \qquad \boldsymbol{\Sigma}_\times=\Sigma_-+i\Sigma_{23};\\
\boldsymbol{\Sigma}_1 &=\Sigma_{12}+i\Sigma_{13}; \qquad \mathbf{v}=v_2+iv_3.
\end{split}
\end{align}
Then the equations of motion for the general tilted type~III/IV/V Bianchi model with diffusion are written as follows:
\begin{align}
\Sigma^\prime_+ &= (q-2)\Sigma_+ +3{\vert \boldsymbol{\Sigma}_1 \vert}^2-2{\vert \mathbf{N}_\times \vert}^2+\frac{\gamma \Omega}{2G_+}(-2v_1^2+{\vert \mathbf{v} \vert}^2), \label{Eq:EqsCons:SigmaPlusEvol}\\
\boldsymbol{\Sigma}^\prime_\times &= (q-2+2i\psi^\prime)\boldsymbol{\Sigma}_\times+\sqrt{3}{\boldsymbol{\Sigma}_1}^2-2\mathbf{N}_\times(iA+\sqrt{3}\bar{N})+\frac{\sqrt{3}\gamma\Omega}{2G_+}\mathbf{v}^2,\\
\boldsymbol{\Sigma}^\prime_1 &= (q-2-3\Sigma_++i\psi^\prime)\boldsymbol{\Sigma}_1-\sqrt{3}\boldsymbol{\Sigma}_\times \boldsymbol{\Sigma}_1^*+\frac{\sqrt{3}\gamma\Omega}{G_+} \mathbf{v}, \\
\mathbf{N}^\prime_\times &= (q+2\Sigma_++2i\psi^\prime)\mathbf{N}_\times+2\sqrt{3}\boldsymbol{\Sigma}_\times \bar{N},\\
\bar{N}^\prime &= (q+2\Sigma_+)\bar{N}+2\sqrt{3}\text{Re}(\boldsymbol{\Sigma}_\times^*\mathbf{N}_\times),\\
A^\prime &= (q+2\Sigma_+)A.
\end{align}
The equations for the fluid are:
\begin{align}
\Omega^\prime &= \frac{\Omega}{G_+} \Biggl \{ 2q-(3\gamma-2)+2\gamma A v_1 + \biggl[ 2q(\gamma-1)-(2-\gamma) - \gamma \mathcal{S} \biggr] V^2 \Biggr \} \nonumber \\
& +\frac{K}{\sqrt{1-V^2}}, \\
\Phi^\prime &= 2(q+1)\Phi-\frac{K}{\sqrt{1-V^2}}, \label{Eq:EqsCons:PhiEvol} \\
v^\prime_1 &= \Biggl[T+2\Sigma_+-\frac{(\gamma-1)\sqrt{1-V^2}G_+}{\gamma G_-} \cdot \frac{K}{\Omega}\Biggr]v_1-2\sqrt{3}\text{Re}(\boldsymbol{\Sigma}_1 \mathbf{v}^*)\nonumber \\
& -A {\vert \mathbf{v} \vert}^2-\sqrt{3}\text{Im}(\mathbf{N}_\times^*\mathbf{v}^2), \label{Eq:EqsCons:v1Evol} \\
\mathbf{v}^\prime &= \Biggl[T-\Sigma_++i\psi^\prime+(A-i\sqrt{3}\bar{N})v_1-\frac{(\gamma-1)\sqrt{1-V^2}G_+}{\gamma G_-} \cdot \frac{K}{\Omega}\Biggr] \mathbf{v}\nonumber \\
& -\sqrt{3}(\boldsymbol{\Sigma}_\times+i\mathbf{N}_\times v_1)\mathbf{v}^*, \label{Eq:EqsCons:vEvol} \\
K^\prime &= \frac{(\gamma-1)G_+V^2}{\gamma \sqrt{1-V^2} G_-}\cdot \frac{K^2}{\Omega}+\Biggl[ 2Av_1+3q-\frac{V^2}{G_-}M \Biggr] K, \label{Eq:EqsCons:DiffEq}
\end{align}
where 
\begin{align}
q &= 2\Sigma^2+\frac{(3\gamma-2)+(2-\gamma)V^2}{2G_+}\Omega-\Phi,\\
V&= \sqrt{v_1^2+{\vert \mathbf{v} \vert}^2}, \\
\Sigma^2 &= \Sigma^2_+ +{\vert\boldsymbol{\Sigma}_\times \vert}^2+ {\vert \boldsymbol{\Sigma}_1\vert}^2, \\
T &= \frac{1}{G_-}\biggl\{ [(3\gamma-4)-2(\gamma-1)Av_1](1-V^2)+(2-\gamma)V^2 \mathcal{S} \biggr\},\\
M &= (3\gamma-4)-2(\gamma-1)Av_1-\mathcal{S},\\
\mathcal{S} &= \frac{1}{V^2}\Sigma_{ab}v^av^b, \\
G_\pm &= 1 \pm (\gamma-1)V^2.
\end{align}
These variables are subject to the following constraints:
\begin{align}
1 &= \Sigma^2+A^2+{\vert \mathbf{N}_\times \vert}^2+\Omega+\Phi,  \label{Eq:EqsCons:HamiltonianC} \\
0 &= 2\Sigma_+A+2\text{Im}(\boldsymbol{\Sigma}_\times^*\mathbf{N}_\times)+\frac{\gamma \Omega v_1}{G_+}, \label{Eq:EqsCons:v1Cons}\\
0 &= \boldsymbol{\Sigma}_1(i\bar{N}-\sqrt{3}A)+i\boldsymbol{\Sigma}_1^*\mathbf{N}_\times+\frac{\gamma \Omega \mathbf{v}}{G_+}. \label{Eq:EqsCons:vCons}
\end{align}

The class of Bianchi type~III=VI$_{-1}$ is given by~${\vert \mathbf{N}_\times \vert}^2-{\bar{N}}^2>0$; type~IV is defined by
${\vert \mathbf{N}_\times \vert}^2-{\bar{N}}^2=0$; and Bianchi type~V cosmologies are given by~$\mathbf{N}_\times=\bar{N}=0$. 
\par 
Function~$\psi$ used in the equations is responsible for the choice of gauge. For type~V cosmologies, we use the gauge freedom to set
\begin{equation}
\text{Im}(\boldsymbol{\Sigma}_1)=\text{Im}(\mathbf{v})=0.
\end{equation} 
For the models of types~III and~IV we choose the so-called 'N-gauge' to set~$N_-=0$ and introduce new geometrical variables~$N$ and~$\lambda$ by
\begin{equation}
N_{23}=N, \quad \bar{N}=\lambda N.
\end{equation}
For type~IV models~$\lambda=\pm 1$ and is therefore a constant, while for type~III cosmologies~$\lambda$ is a function restricted by~$\vert \lambda \vert <1$. In addition, this function obeys
\begin{align}
\lambda^\prime &= 2\sqrt{3}\Sigma_{23}(1-\lambda^2),\label{Eq:EqsCons:LambdaEvol}\\
0 &=					A^2-3(1-\lambda^2)N^2. \label{Eq:EqsCons:LambdaCons}
\end{align}

The state vector of a Bianchi type III~model is by this
\begin{equation}
\mathbf{X}=\Bigl[ \Sigma_+, \boldsymbol{\Sigma}_\times, \boldsymbol{\Sigma}_1, N, \lambda, A, \Omega, \Phi, v_1, \mathbf{v}, K  \Bigr]
\end{equation}
modulo the constraints~(\ref{Eq:EqsCons:HamiltonianC})--(\ref{Eq:EqsCons:vCons}),~(\ref{Eq:EqsCons:LambdaCons}). For type~IV we exclude~$\lambda$ from the list of variables and equation~(\ref{Eq:EqsCons:LambdaCons}) from the list of constraints. So, the evolution takes place on a 9-dimensional subspace of a 14- and 13-dimensional manifold for type~III and type~IV cosmologies, respectively. In both cases, the physical state space is 9-dimensional.
\par
The state vector describing the Bianchi type~V cosmological model is
\begin{equation}
\mathbf{X}=\Bigl[ \Sigma_+, \Sigma_-, \Sigma_{12}, \Sigma_{23}, A, \Omega, \Phi, v_1, v_2, K  \Bigr]
\end{equation}
modulo the constraints (\ref{Eq:EqsCons:HamiltonianC})--(\ref{Eq:EqsCons:v1Cons}). Therefore, the evolution takes place on a 7-dimensional subspace of a 10-dimensional manifold; the dimension of the physical state space is seven.
\par 
Thus, taking diffusion into account results in certain changes in the equations for the fluid; namely, the evolution equations for the energy density~$\Omega$, the scalar potential~$\Phi$, and the tilt variables~$v_a$ are modified by presence of extra terms. Also, the system gets an additional equation arising from the current density conservation~(\ref{Eq:Model:Conservation}). The original diffusion-free equations can be retained by taking~$K$ to be identically zero.

\par 
A known important feature of the diffusionless equations describing the Bianchi cosmologies of considered types is that the state space is compact (see e. g. \cite{Christiansen2008}). In the absence of diffusion the equation~(\ref{Eq:EqsCons:PhiEvol}) grants the non-negativity of the scalar potential~$\Phi$, and the variables are bounded by the Hamiltonian constraint~(\ref{Eq:EqsCons:HamiltonianC}). However, the variable~$K$ appearing in diffusive models has no upper bound; the only restriction on~$K$ is its non-negativity~($K\geq 0$). Due to the additional term in the evolution equation~(\ref{Eq:EqsCons:PhiEvol}), the scalar potential can change its sign and decrease without bound, provided~$K$ is sufficiently large.  This automatically removes the restriction on the other variables, yielding a non-compact state space. 

\subsection{Initial conditions}
\label{Sec:EqsCons:IC}

The initial conditions are chosen so that all the constraints are satisfied at~$\tau=0$. We have consider a wide variety of different sets of initial values of the variables. However, we are particularly interested in investigating the dynamics of cosmologies being initially close to the flat Robertson-Walker model, for which
\begin{align}
\begin{split}
\Sigma_{ab} &=0; \quad N_{ab}=0; \quad A=0;\\
\Omega &= 1; \quad \Phi=0; \quad V=0; \quad K=0; \\
\lambda &= \lambda_0~\text{(for type III models)}. \label{Eq:IC:RW}
\end{split}
\end{align}
So the tables and plots in the given work are presented for the following set of values close to the values~(\ref{Eq:IC:RW}):
\begin{align}
\begin{split}
\Sigma_- & = \Sigma_{23}=0.1; \quad \Sigma_+ = -0.05; \\
\Sigma_{12} &= 0.01; \quad A=0.1; \\
\text{(types III and IV)} \quad \Sigma_{13} &= 0.01; \quad N=0.1;\\  
\Omega &= 0.9. \label{Eq:IC:Inconds}
\end{split}
\end{align} 
Regarding the initial value~$K_0$ of the diffusion term~$K$, to investigate the diffusional effects we consider different values from the interval
\begin{equation}
0\leq K_0 <0.12.
\end{equation}
Since the Bianchi models possess certain symmetric properties, we can choose the positive initial values for~$\lambda$ without loss of generality:~$\lambda=1$ for type~IV and~$\lambda=\sqrt{2/3}$ for type~III. The initial value of~$\Phi$ is determined by the Hamiltonian constraint~(\ref{Eq:EqsCons:HamiltonianC}). The fluid velocity components can then be found by solving the constraints~(\ref{Eq:EqsCons:v1Cons})--(\ref{Eq:EqsCons:vCons}) algebraically.

\section{Results}
\label{Sec:Results}
For every case considered the numerical integration of the complete system has been done over a wide time range~$(-40<\tau<25)$. Note however that shown in the figures are relatively narrow intervals of the time axis. This is made for illustrative purposes.
\par 
Due to the feature that type~IV models demonstrate the same qualitative and nearly the same quantitative behaviour as type~III universes, most results are shown for type~III cosmologies. Unless stating otherwise, the descripton of dynamics of type~III models is also valid for type~IV. The most essential difference between these models is encapsulated in the specific~$\lambda$-variable, see section~\ref{Sec:Results:Lambda}.

\subsection{Future attractors and possible recollapse}
\label{Sec:Results:Recollapse}

In the absence of diffusion the scalar potential represents "ordinary" cosmological constant, which ultimately dominates the evolution. At later times the models enter the accelerated (namely, exponential) expansion stage, and the geometries asymptotically approach the de Sitter state:
\begin{align}
\begin{split}
\label{Eq:Results:deSitter}
\Sigma_{ab} &=0; \quad A=0; \quad N=0;\\
\Omega &= 0; \quad \Phi=1.
\end{split}
\end{align}
Regarding the tilt, only fluids with~$\gamma<4/3$ isotropize, but ones stiffer than radiation do not. So, in the absence of diffusion we observe~$V\to 0$ for dust and~$V\to V^*$ for radiation, where~$0<V^*<1$. Such behaviour is also predicted teoretically by the cosmic no-hair theorem~\cite{Wald1983,Coley2005,Hervik2006,Christiansen2008}.
\par 
The situation is different if one allows for a positive initial value of the diffusion term~$K$. The no-hair theorem is no longer applicable because of the "self-interacting" energy-momentum tensor~(\ref{Eq:Model:CompositeTensor}).
The numerical simulations reveal that diffusive models are ever-expanding only if the diffusion term is initially small enough; higher values of~$K_0$ cause the universes to recollapse (the same has been discovered before for~FRW,~$G_2$ and Bianchi type~VIII cosmologies, see~\cite{Calogero2011},~\cite{Shogin2013}, and~\cite{Shogin2014a} respectively). If the initial conditions for all the variables excluding~$K$ are fixed, the "critical" value~$K_r$ (the minimal value of~$K_0$ at which the universe recollapses), is determined by the model type and the kind of the fluid.

\par 
The values of~$K_r$, and also the timepoints~$\tau_r$ of recollapse (the moments when the dimensionless variables run into singularity) are given in Table~\ref{Tab:Criticals}. It can be seen from the table that dust-filled models are less sensitive to diffusion. Namely, in the case of dust recollapse arrives at later times, for example for Bianchi type~III cosmologies we observe~$\tau_r=4.58$ for radiation against~$\tau_r=5.13$ for dust. Also, the critical initial values~$K_r$ of the diffusion term for dust are approximately~20\% higher than those for radiation. This difference can be expected from an examination of the system of equations in section~\ref{Sec:EqsCons:Eqs}. Setting~$\gamma=1$ eliminates the diffusional parts in equations~(\ref{Eq:EqsCons:v1Evol})--(\ref{Eq:EqsCons:vEvol}); also, one of the terms in the diffusion equation~(\ref{Eq:EqsCons:DiffEq}) vanishes. So, one can say that the dust-filled models might be more stable with respect to the diffusion term variations in comparison to the radiation-filled cosmologies.

\begin{table}[ht]
\centering
\begin{tabular}{|c|c|c|c|c|c|c|}
\hline	
Cosmology		&	\multicolumn{2}{c|}{Type III} 	&	\multicolumn{2}{c|}{Type IV}		&	\multicolumn{2}{c|}{Type V} \\
\hline 
$\gamma$			&	1	&	4/3	&	1	&	4/3	&	1	&	4/3	\\
\hline 
$K_r$				&	0.09538	&	0.08072	&	0.09539	&	0.08073	&	0.11291	&	0.09584 \\
\hline 
$\tau_r$			&	5.13	&	4.58	&	5.28	&	5.04	&	4.99	&	4.52	\\
\hline 
\end{tabular}
\caption{Critical values of the diffusion term~$K_r$ and timepoints of recollapse~$\tau_r$ for Bianchi cosmologies of types~III,~IV, and~V, filled with dust and radiation}
\label{Tab:Criticals}
\end{table}

\par 
Since the quantitative behaviour of Bianchi type~III and~IV cosmologies with diffusion is in fact nearly the same, so are the values of~$K_r$ for these models (see Table~\ref{Tab:Criticals}). However, type~IV models are observed to recollapse at later times, which possibly can be explained by different nature of the~$\lambda$-variable.

\par 
Henceforth, we will restrict our consideration to the cases with~$K_0<K_r$ only, since ever-expanding models are of particular interest. Furthermore, recollapse can occur only if the scalar potential becomes negative, which has no clear physical meaning (albeit a mathematical possibility).

\subsection{The future attractor for ever-expanding models}
We start with assuming that in an ever-expanding model, the scalar potential asymptotically behaves as a cosmological constant, namely
\begin{equation}
\Phi \to 1~\text{as}~\tau\to\infty.
\end{equation}
This conjecture is confirmed by numerical simulations. It then follows from the Hamiltonian constraint,~(\ref{Eq:EqsCons:HamiltonianC}), that
\begin{equation}
[\Sigma_{ab}, A, N_{ab}, \Omega] \to [0,0,0,1],~\text{as}~\tau\to \infty.
\end{equation}
In Bianchi type~III models, the equation~(\ref{Eq:EqsCons:LambdaEvol}) reduces then to~$\lambda^\prime=0$, describing the fact that~$\lambda$ approaches a constant at late times. The remaining evolution equations for the tilt components,~(\ref{Eq:EqsCons:v1Evol})-(\ref{Eq:EqsCons:vEvol}), and the diffusion term,~(\ref{Eq:EqsCons:DiffEq}), are undefined on the vacuum boundary~($\Omega=0$) and in case of extreme tilt~($V=1$). As in~\cite{Shogin2014a}, we introduce a new variable~$Y$ to eliminate this problem:
\begin{equation}
Y=\frac{K}{\Omega\sqrt{1-V^2}}.
\end{equation}
The corresponding evolution equation is derived from the diffusion equation~(\ref{Eq:EqsCons:DiffEq}). The late-time evolution of the models is thus determined by the future attractors of the two-dimensional system of differential equations for~$V$ and~$Y$. These attractors are in fact similar to those obtained in~\cite{Shogin2014a} for Bianchi type~VIII cosmologies: in the case of an arbitrary~$\gamma$-fluid
\begin{align}
\begin{split}
\label{Eq:Results:Attractor}
0<\gamma\leq 1: & \quad \text{at}~\tau\to \infty~[V,Y]\to[0,0],\\
1<\gamma<3/2: 	 & \quad \text{at}~\tau\to \infty~[V,Y]\to[0,3(\gamma-1)],\\
\gamma=3/2:		 & \quad \text{at}~\tau\to \infty~[V,Y]\to[\bar{V},3/(2+\bar{V}^2)],\\
3/2<\gamma<2: 	 & \quad \text{at}~\tau\to \infty~[V,Y]\to[1,1].
\end{split}
\end{align}
Note that~(\ref{Eq:Results:Attractor}) implies~$K\to 0$ at~$\tau\to \infty$. Also, it can be seen that~$\gamma$ has two bifurcation values, namely~$\gamma=1$ and~$\gamma=3/2$. The latter can be referred to as the tilt bifurcation, since it switches on the asymptotic tilt. For the models without diffusion, as mentioned in section~\ref{Sec:Results:Recollapse}, the tilt bifurcation occurs at a lower value of~$\gamma$~($\gamma=4/3$).

\subsection{Asymptotic behaviour of the tilt}

\begin{figure}[t]
\begin{minipage}[h]{0.49\linewidth}
\includegraphics[width=1.0\linewidth]{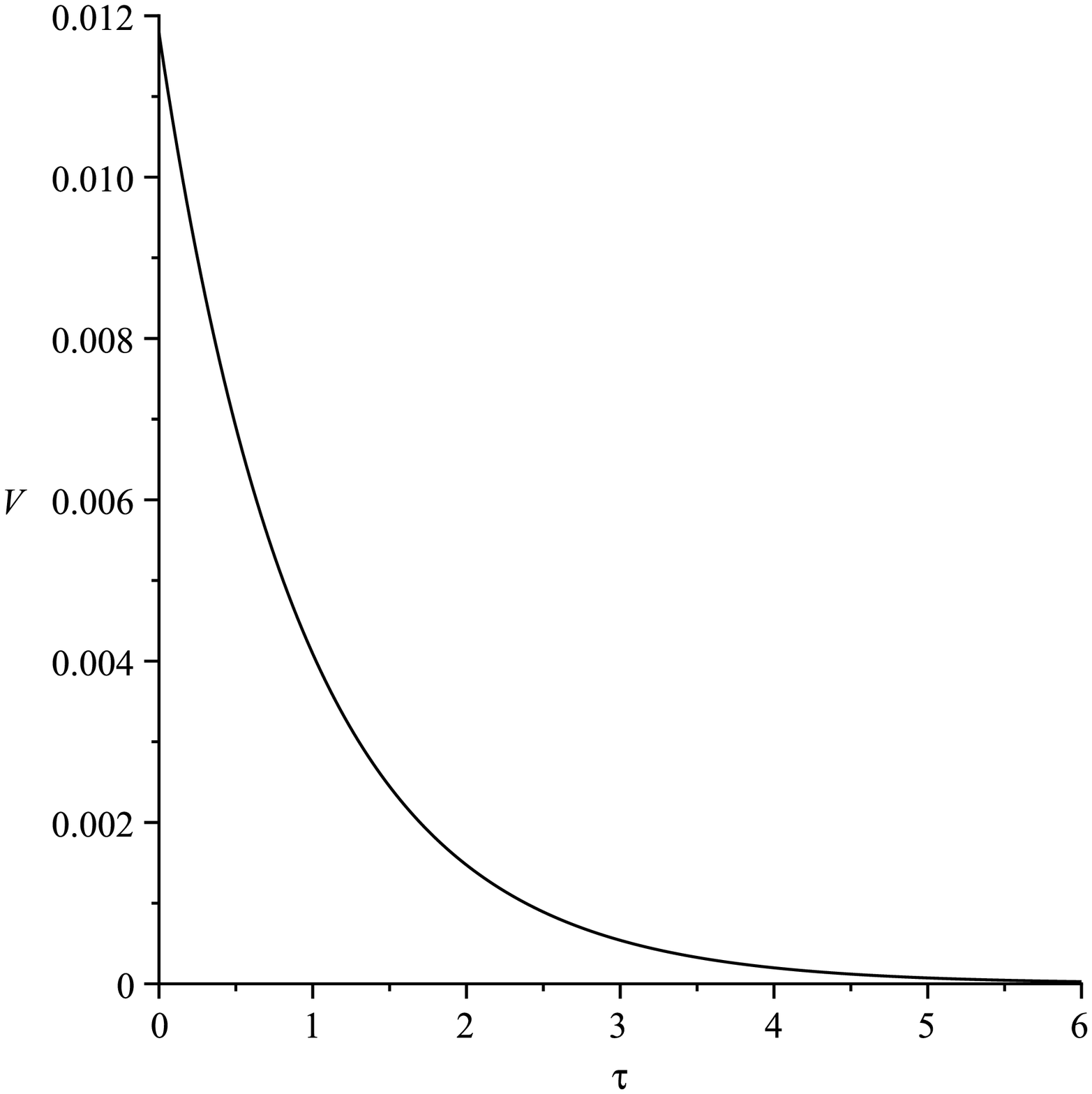}
\end{minipage}
\begin{minipage}[h]{0.49\linewidth}
\includegraphics[width=1.0\linewidth]{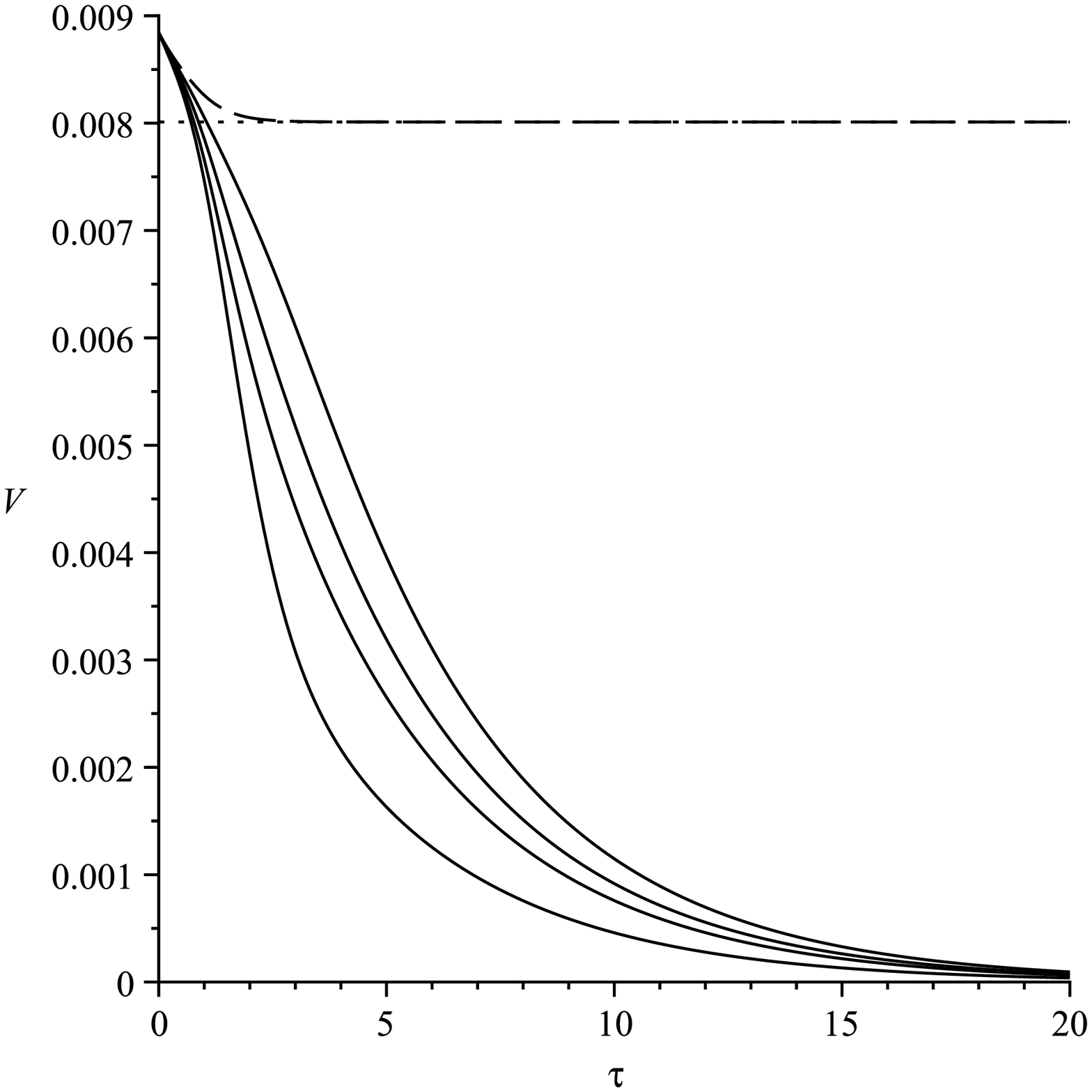}
\end{minipage}
\caption{Dynamics of the tilt magnitude~$V$ for Bianchi type III models filled with dust at~${0\leq K_0<K_r}$ (left), and radiation at~$K_0=0; \, 0.02; \, 0.04; \, 0.06; \, 0.08$ (right). Note that in the radiation case the fluid is asymptotically tilted in the absence of diffusion, but becomes non-tilted when diffusion is present}
\label{Fig:Velocities}
\end{figure}

For the dust-filled cosmologies, the diffusion impact on the fluid velocity dynamics is negligibly small. In fact, the tilt is slightly reduced at higher values of~$K_0$, but the reduction does not exceed~1\% even in the cases when diffusion is extremely strong~($K_0\to K_r$). At sufficiently late times the tilt of dust decays exponentially:~$V\sim e^{-\tau}$. The dynamics of the tilt in the dust-filled Bianchi III model can serve as a typical example, see Figure~\ref{Fig:Velocities}, left.

\par 
The situation changes drastically for the cosmologies filled with radiation. Since in the diffusionless case~$\gamma=4/3$ corresponds to the tilt bifurcation value, the tilt components freeze into some small constant values in such models. On the contrary, in presence of diffusion the fluid {\it does} isotropize. A closer investigation reveals that this happens exponentially; namely, at sufficiently late times $v_1,v_2,v_3\sim e^{-0.25\tau}$, which is in agreement with results obtained in~\cite{Shogin2014a} for a Bianchi type~VIII model.
\par 
Moreover, the higher the value~$K_0$ is, the smaller the proportionality constants are; that is, stronger diffusion causes more significant reduction of the tilt.
\par 
As an example, the tilt evolution in radiation-filled Bianchi type~III universes is demonstrated in Figure~\ref{Fig:Velocities}, right. The solid lines in the figure correspond to different initial values of~$K_0=0.02;\,0.04;\,0.06;\,0.08$~(the higher~$K_0$, the lower the curve). The case without diffusion is shown by the dashed line.

\subsection{Future asymptotic behaviour of energy density and scalar potential}
\label{Sec:Results:OmegaPhi}
The simulation results show that presence of diffusion changes the future asymptotic form for the energy density~$\Omega$ in ever-expanding universe models. More precisely, in the diffusive case~$\Omega$ is decreasing slower than in the absence of diffusion.

\par 
For the radiation-filled models, at sufficiently late times
\begin{equation}
\Omega \sim \left\{ \begin{array}{ll} 
e^{-4\tau}, & \text{without diffusion}; \\
e^{-3\tau}, & \text{with diffusion}.
\end{array} \right.
\end{equation}
The situation is more complicated for cosmologies filled with dust. The leading terms can be written as
\begin{equation}
\left\{ \begin{array}{lll} 
\Omega \sim & e^{-3\tau}, & \text{without diffusion}; \\
\Omega \to  & (C_K\cdot\tau+C_\Omega)e^{-3\tau}, & \text{with diffusion}.
\end{array} \right.
\end{equation}
Here~$C_K,~C_\Omega$ are two constants; moreover,~$C_K$ is actually the proportionality constant in the asymptotic expression for the diffusion term~$K$, see section~\ref{Sec:Results:Diffusion}, later. Both terms play a substantial role, since~$C_\Omega$ is found to be sufficiently large. For example, numerical investigation of a dust-filled type~V model yields~${C_K\approx 2.60,}$ ${C_\Omega\approx 15.82}$ at~${K_0=0.03}$, and~${C_K\approx 14.64,}$ ${C_\Omega\approx 18.71}$ at~${K_0=0.07}$.
\par 
The statement that the energy density of radiation decreases faster than that of dust (see e.\,g.~\cite{Groen2007}) holds also in the diffusive case. However, the ratio~$\Omega_{rad}/\Omega_{dust}$ decays significantly faster when diffusion is absent.

\begin{figure}[ht]
\begin{minipage}[h]{0.49\linewidth}
\includegraphics[width=1.0\linewidth]{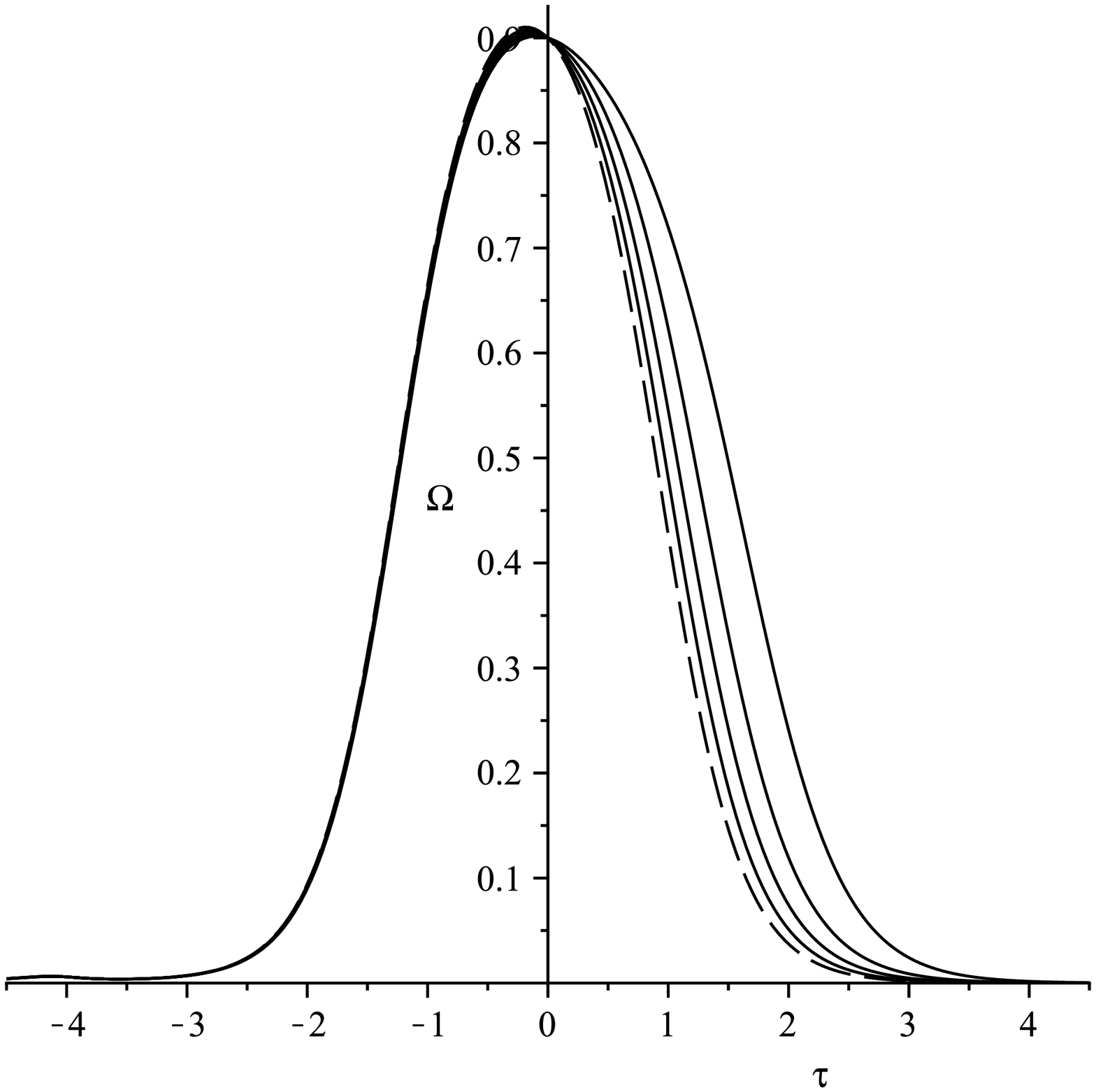}
\end{minipage}
\begin{minipage}[h]{0.49\linewidth}
\includegraphics[width=1.0\linewidth]{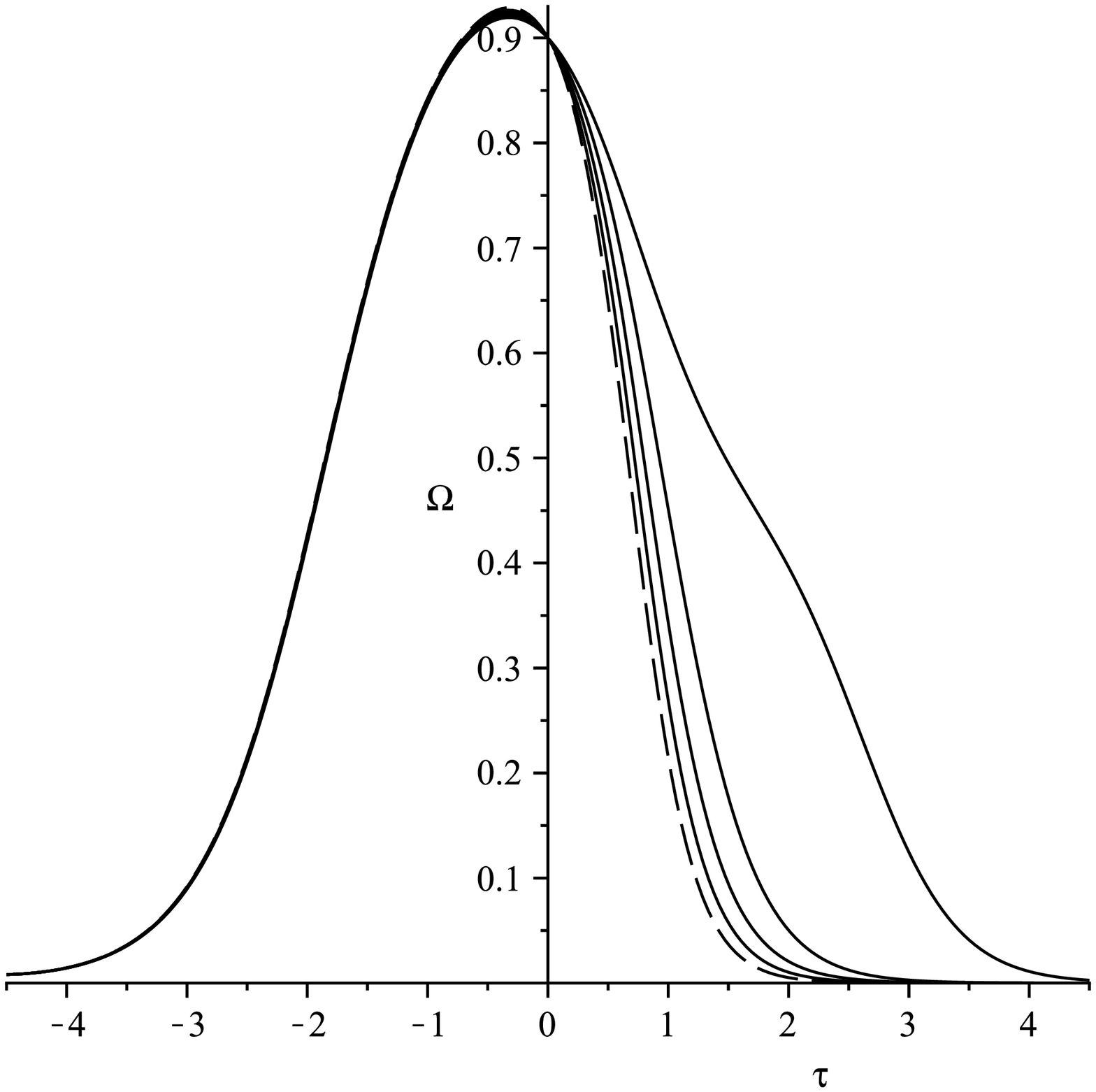}
\end{minipage}
\caption{Dynamics of the energy density~$\Omega$ for Bianchi type~III models filled with dust (left) and radiation (right) for~$K_0=0; \, 0.02; \, 0.04; \, 0.06; \, 0.08$}
\label{Fig:Omegas}
\end{figure}

The influence of diffusion on the energy density is demonstrated in Figure~\ref{Fig:Omegas}, where Bianchi type~III models are taken as an example.  
The solid lines show the dynamics of~$\Omega$ at different values of $K_0=0.02; \, 0.04; \, 0.06; \, 0.08$ (higher curves correspond to larger values of~$K_0$). The dashed lines denote the diffusionless case. There is a significant deformation of the plot in the case~$\gamma=4/3,~K_0=0.08,$ which is explained by the proximity of~$K_0$ to the critical value for this model~$K_r\approx 0.0807.$

\par
\begin{figure}[ht]
\begin{minipage}[h]{0.49\linewidth}
\includegraphics[width=1.0\linewidth]{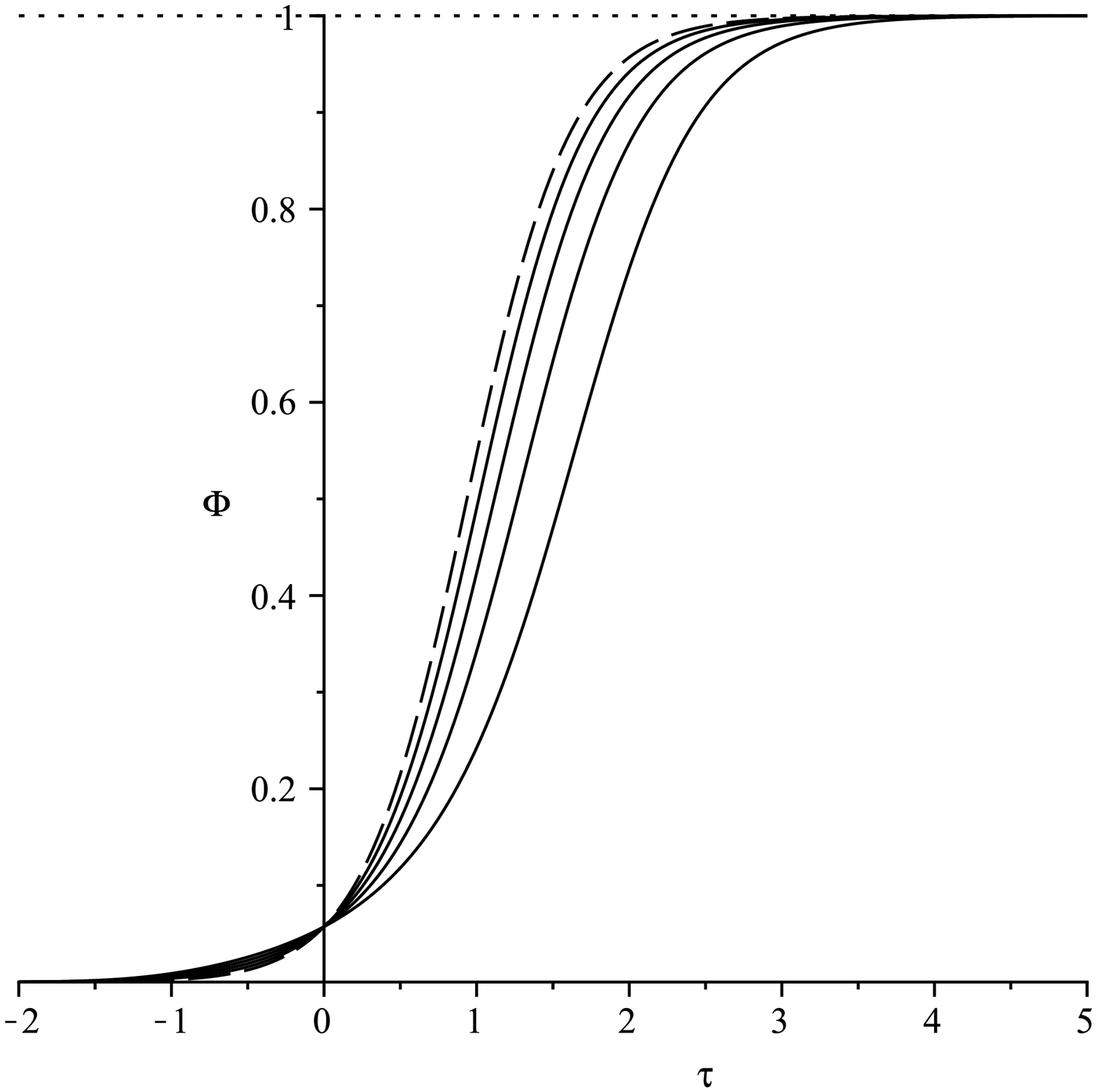}
\end{minipage}
\begin{minipage}[h]{0.49\linewidth}
\includegraphics[width=1.0\linewidth]{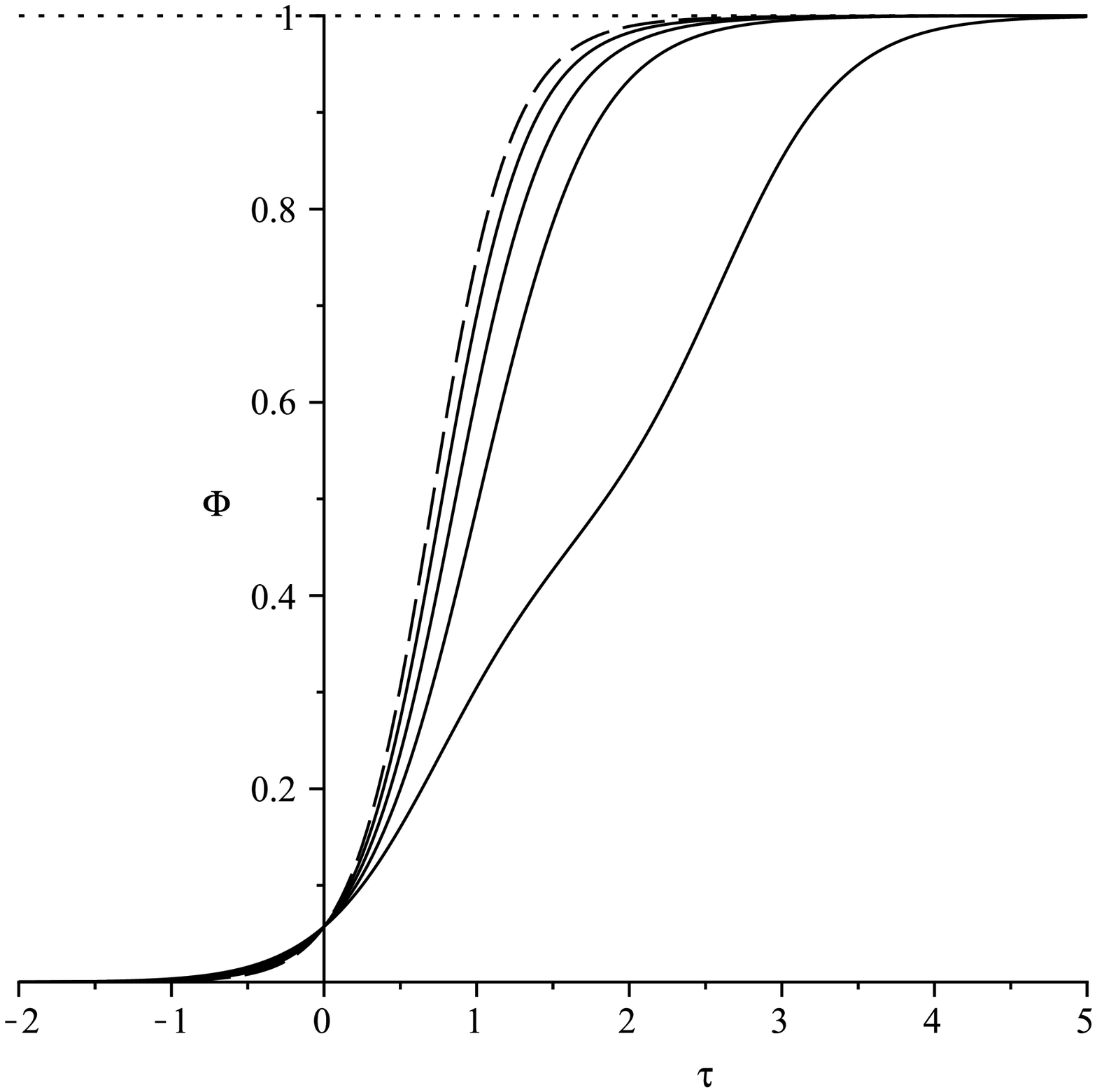}
\end{minipage}
\caption{Dynamics of the scalar potential~$\Phi$ for Bianchi type~III models filled with dust (left) and radiation (right) for~$K_0=0; \, 0.02; \, 0.04; \, 0.06; \, 0.08$}
\label{Fig:Phis}
\end{figure}

 
On the other hand, the future asymptotic form for the scalar potential is described by
\begin{equation}
1-\Phi \sim e^{-2\tau},
\end{equation}
both for dust and radiation and is in fact independent of diffusion. The only effect on the scalar potential is quantitative; namely, the stronger the diffusion is, the less rapidly the potential grows. This can be seen in Figure~\ref{Fig:Phis}, where the dynamics of~$\Phi$ is given for type~III models filled with dust and radiation. The solid lines correspond to different initial values of ${K_0=0.02; \, 0.04; \, 0.06; \, 0.08}$ (the larger~$K_0$ is, the slower the potential grows). The dynamics in the diffusionless case is shown by the dashed lines. The deviations from the diffusionless dynamics are small at lesser values of~$K_0$ but become more significant as~$K_0$ comes closer to~$K_r$. The diffusional effect is most drastic in the case~$\gamma=4/3,~K_0=0.08,$ which is close to the critical value for the given model~$(K_r\approx 0.0807).$

\subsection{Diffusion term dynamics}
\label{Sec:Results:Diffusion}
For non-recollapsing Bianchi models the diffusion term~$K$ is asymptotically zero in the past and first grows monotoneously. It passes~$K_0$ at~$\tau=0$ and  reaches its maximal value at some positive time~$\tau_K$. After that,~$K$ starts to decrease monotoneously and tends exponentially to zero in the future: similarly to Bianchi type~VIII~\cite{Shogin2014a}, at sufficiently late times~$K\sim e^{-3\tau}$ both for dust- and radiation-filled models of considered types.

\par 
One can introduce two characteristical timepoints to mark the beginning and the end of the stage of cosmological evolution where the diffusion term is significant. Then, for all the considered models increasing~$K_0$ leads to the following effects: the starting point for the diffusion stage is shifted to the left and the end point to the right, the latter being much more significant; the maximal value of the diffusion term becomes reached at later times and is higher at larger values of~$K_0$.

\begin{figure}[ht]
\begin{minipage}[h]{0.49\linewidth}
\includegraphics[width=1.0\linewidth]{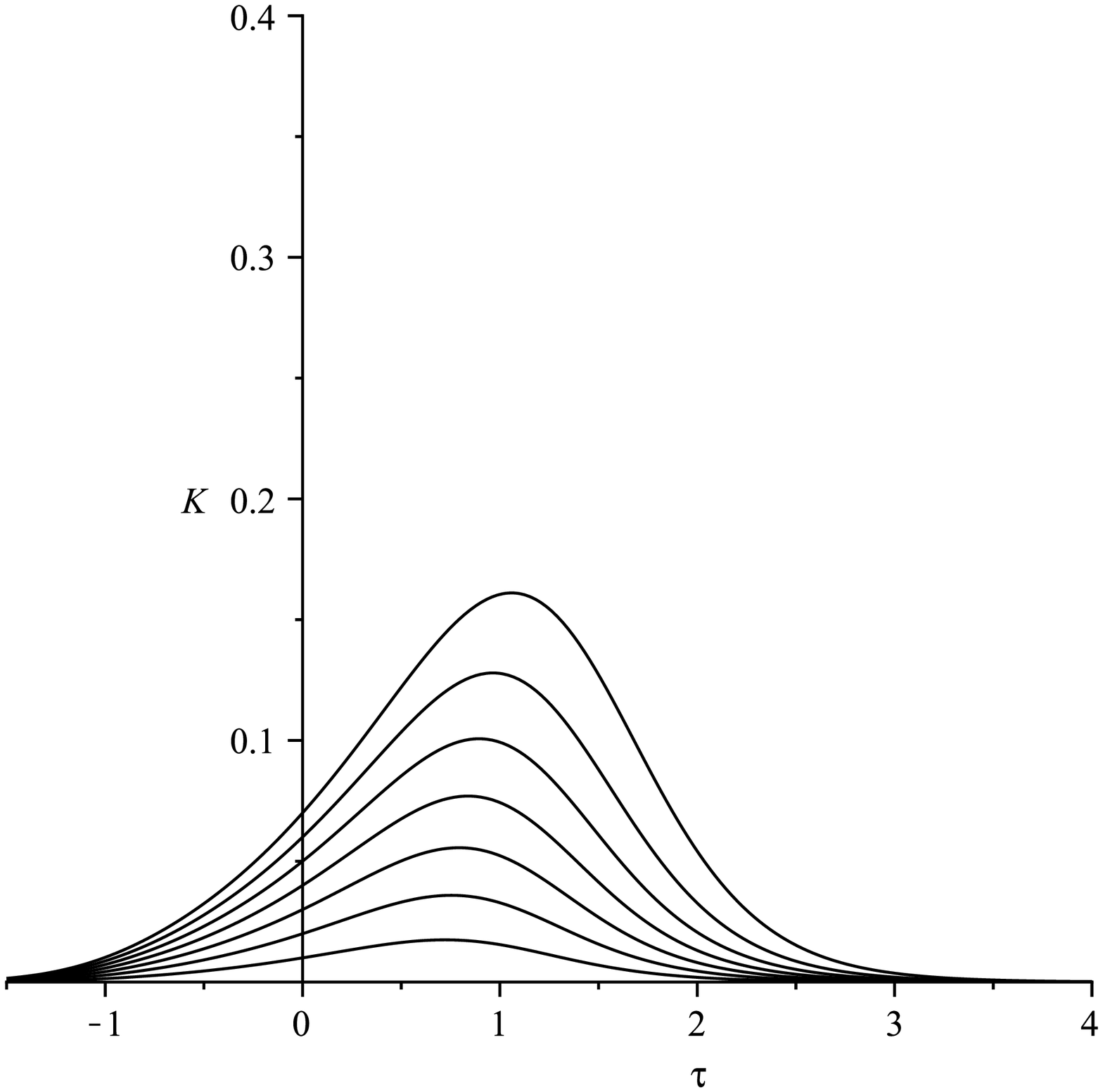}
\end{minipage}
\begin{minipage}[h]{0.49\linewidth}
\includegraphics[width=1.0\linewidth]{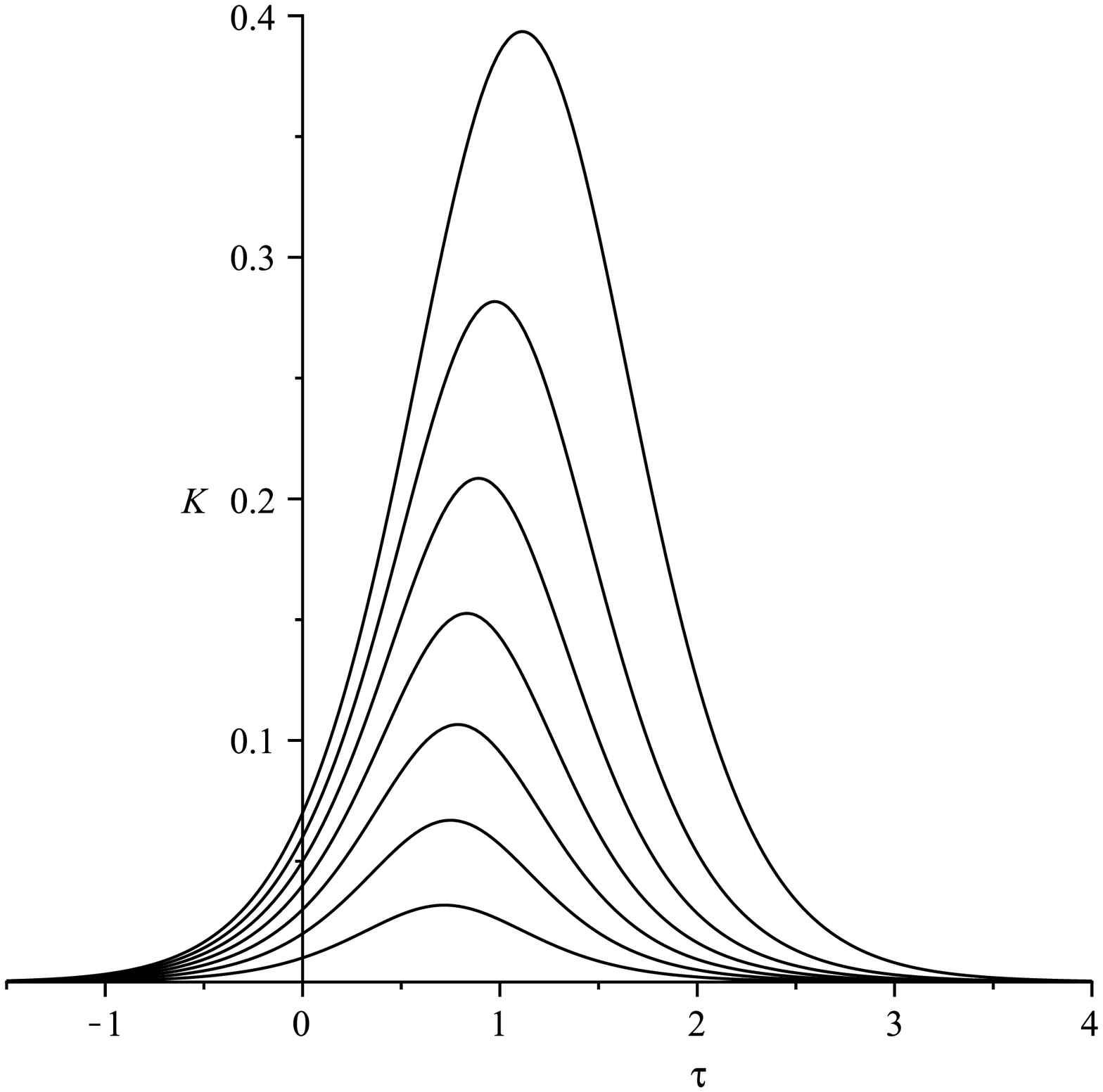}
\end{minipage}
\caption{Dynamics of the diffusion term~$K$ for Bianchi type~III models at~$\gamma=1$ (left) and~$\gamma=4/3$ (right) for~$K_0=0.01\dots 0.07$}
\label{Fig:Ks}
\end{figure}

\begin{table}[ht]
\centering
\begin{tabular}{|c|c|c|c|c|c|c|c|c|c|c|}
\hline 
Quantity			& \multicolumn{4}{c|}{$\tau_K$} & \multicolumn{4}{c|}{$K_{max}$} \\
\hline	
Cosmology		&	\multicolumn{2}{c|}{Type III} 	&		\multicolumn{2}{c|}{Type V} & \multicolumn{2}{c|}{Type III} & \multicolumn{2}{c|}{Type V} \\
\hline 
$\gamma$			&	1			&	4/3		&	1			&	4/3		&	1			&	4/3	&	1			&	4/3\\
\hline 
$K_0=0.02$		&	0.75		&	0.75		&	0.69		&	0.70		&	0.036		&	0.067		&	0.036		&	0.060	\\
\hline 
$K_0=0.04$		&	0.84		&	0.83		&	0.75		&	0.76		&	0.077		&	0.153		&	0.071		&	0.133	\\
\hline
$K_0=0.06$		&	0.96		&	0.97		&	0.84		&	0.85		&	0.128		&	0.282		&	0.113		&	0.229	\\
\hline 
$K_0=0.08$		&	1.20		&	1.69		&	0.97		&	1.02		&	0.206		&	0.723		&	0.168		&	0.385	\\
\hline
\end{tabular}
\caption{The time of maximum,~$\tau_K,$ and the corresponding maximal value~$K_{max}$ of the diffusion term as functions of~$K_0$ for Bianchi cosmologies of types~III and~V filled with dust and radiation}
\label{Tab:K}
\end{table}

\par 
The dynamics of the diffusion term~$K$ for type~III universes filled with dust and radiation is shown in Figure~\ref{Fig:Ks}. The curves are drawn for the different values of~$K_0$ from~0.01 to~0.07 with a step of~0.01. The diffusion term reaches greater values in the case of radiation, corresponding to the prediction that dust should be less sensitive to the diffusional effects. Also shifting of~$\tau_K$ to the right with growth of~$K_0$ can be seen in the figure.
\par 
Some numerical results can be found in Table~\ref{Tab:K}. It can be seen from the table that under the same conditions the diffusion term takes lesser values in type~V cosmologies than in type~III. The difference becomes most noticeable at higher values of~$K_0$: for example, at~${K_0=0.08}$ the maximal value of the diffusion term for type~III cosmologies is approximately~23\% and~87\% higher for dust and radiation, respectively, than the corresponding values for Bianchi type~V universes. 
\par 
One can also see from Table~\ref{Tab:K} that for the same model type, the position of timepoint~$\tau_K$ in fact does not depend on the kind of the fluid when diffusion is weak enough, while at higher values of~$K_0$ this characteristical timepoint arrives later for dust. This is explained by the fact that at the same relatively high value of~$K_0$ the radiation-filled model is closer to its critical point~$K_r$, since the values of~$K_r$ for the dust are higher than for radiation, as mentioned in section~\ref{Sec:Results:Recollapse}.

\subsection{Quantitative impact on geometrical variables}
\label{Sec:Results:Quantitative}
From a physical point of view, presence of diffusion enables energy transfer between the two particle systems; namely, in our case energy is transferred from the scalar field to the matter and thus slows down the process of cosmological evolution~\cite{Calogero2011,Shogin2013}. As we have shown in section~{\ref{Sec:Results:OmegaPhi}, the additional energy changes the asymptotical behaviour of energy density and makes this quantity decrease in a slower manner than without diffusion. Although the late-time behaviour of the geometrical variables is not affected qualitatively, this leads to some typical quantitative effects. The more intense the diffusion is, the stronger these effects become.

\par 
For any particular moment in time~$\tau>0$ the energy density of the fluid is higher if the diffusion is present, the other conditions being the same. This also holds for the geometrical variables, that is
\begin{equation}
\vert Y_G (K_{02})\vert > \vert Y_G(K_{01})\vert
\end{equation}
at a fixed time~$\tau>0$, where~$Y_G$ denotes a geometrical variable, and~$K_{02}>K_{01}$ are some certain initial values of the diffusion term. In particular, the maxima of~$\vert Y_G(\tau) \vert$, if any, become more significant as diffusion grows stronger. 
\par 
If one introduces some characteristical timepoints describing the evolution of a geomertical variable~$Y_G$, these points are shifted to the right under conditions of stronger diffusion. Particularly, the maxima and minima of~$\vert Y_G(\tau) \vert$, if observed, will appear at later times.

\begin{figure}[ht]
\begin{minipage}[h]{0.49\linewidth}
\includegraphics[width=1.0\linewidth]{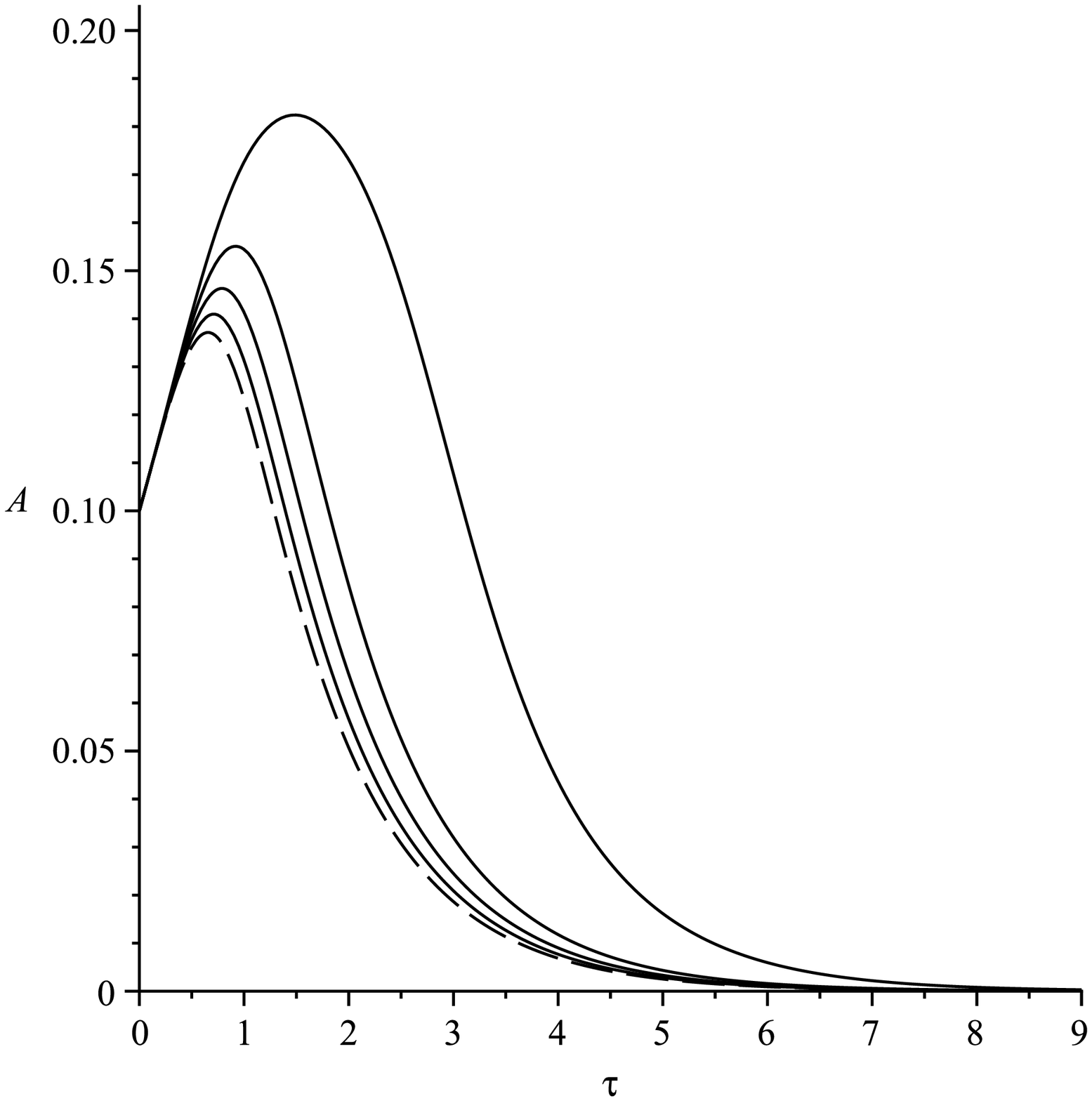}
\end{minipage}
\begin{minipage}[h]{0.49\linewidth}
\includegraphics[width=1.0\linewidth]{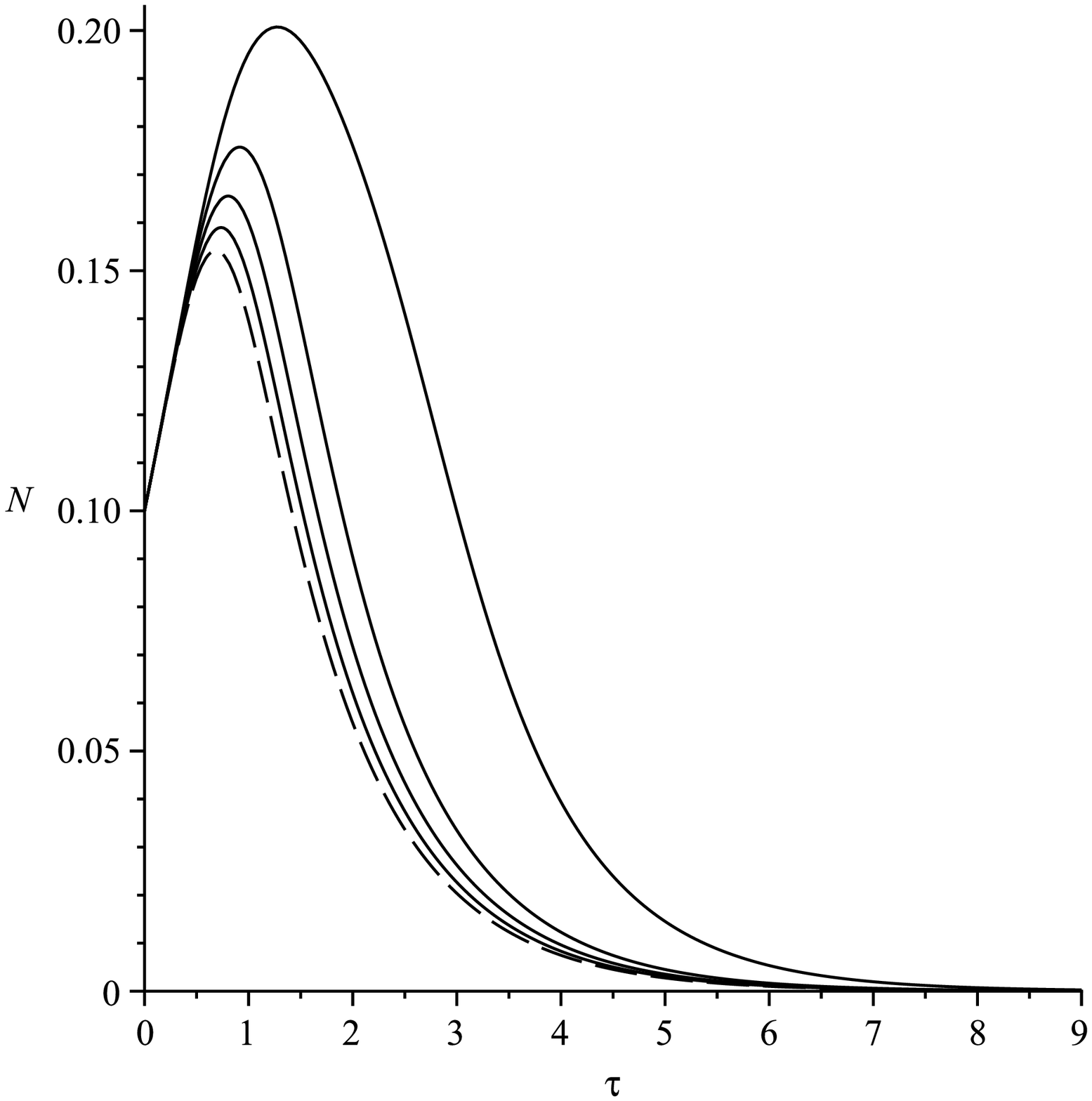}
\end{minipage}
\caption{Dynamics of the curvature variables~$A$~(left) and~$N$~(right) for Bianchi type~III cosmologies at~$\gamma=4/3$ for~$K_0=0;\,0.02;\,0.04;\,0.06;\,0.08$}
\label{Fig:AN}
\end{figure}

\par 
To visualise these effects, we demonstrate the dynamics of the curvature variables~$A$ and~$N$ for radiation-filled Bianchi type~III universes at different initial values of diffusion term, see Figure~\ref{Fig:AN}. The solid lines show the evolution of variables at~$K_0=0.02;\,0.04;\,0.06;\,0.08;$ higher curves correspond to greater values of~$K_0$. The diffusionless case is depicted by the dashed lines. It can be seen from the plots that the maxima which are observed in the evolution of the given quantities become stronger at greater values of~$K_0$, the points of maximum being at the same time shifted to the right. Corresponding numerical data can be found in Tables~\ref{Tab:A} and~\ref{Tab:N}.

\begin{table}[ht]
\centering
\begin{tabular}{|c|c|c|c|c|c|c|c|c|c|c|}
\hline 
Quantity			& \multicolumn{4}{c|}{$\tau_A$} & \multicolumn{4}{c|}{$A_{max}$} \\
\hline	
Cosmology		&	\multicolumn{2}{c|}{Type III} 	&		\multicolumn{2}{c|}{Type V} & \multicolumn{2}{c|}{Type III} & \multicolumn{2}{c|}{Type V} \\
\hline 
$\gamma$			&	1		&	4/3	&	1		&	4/3	&	1		&	4/3	&	1		&	4/3	\\
\hline 
$K_0=0	$		&	0.63	&	0.66	&	0.60	&	0.63	&	0.114	&	0.137	&	0.113	&	0.135	\\
\hline 
$K_0=0.02$		&	0.70	&	0.71	&	0.65	&	0.68	&	0.115	&	0.141	&	0.114	&	0.138	\\
\hline 
$K_0=0.04$		&	0.78	&	0.79	&	0.72	&	0.74	&	0.118	&	0.146	&	0.116	&	0.142	\\
\hline
$K_0=0.06$		&	0.90	&	0.92	&	0.81	&	0.83	&	0.121	&	0.155	&	0.118	&	0.148	\\
\hline 
$K_0=0.08$		&	1.13	&	1.48	&	0.95	&	1.00	&	0.127	&	0.182	&	0.122	&	0.160	\\
\hline
\end{tabular}
\caption{The time of maximum, $\tau_A,$ and the corresponding maximal value~$A_{max}$ of the curvature variable~$A$ for Bianchi type~III and type~V cosmologies filled with dust and radiation}
\label{Tab:A}
\end{table}

\begin{table}[ht]
\centering
\begin{tabular}{|c|c|c|c|c|c|c|}
\hline	
Quantity		&	\multicolumn{2}{c|}{$\tau_N$} 	&		\multicolumn{2}{c|}{$N_{max}$} \\
\hline 
$\gamma$			&	1		&	4/3	&	1		&	4/3	\\
\hline 
$K_0=0	$		&	0.69	&	0.69	&	0.127	&	0.154	\\
\hline 
$K_0=0.02$		&	0.74	&	0.74	&	0.129	&	0.159	\\
\hline 
$K_0=0.04$		&	0.80	&	0.80	&	0.132	&	0.166	\\
\hline
$K_0=0.06$		&	0.91	&	0.91	&	0.136	&	0.176	\\
\hline 
$K_0=0.08$		&	1.09	&	1.28	&	0.143	&	0.201	\\
\hline
\end{tabular}
\caption{The time of maximum, $\tau_N,$ and the corresponding maximal value~$N_{max}$ of the curvature variable~$N$ for Bianchi type~III models filled with dust and radiation}
\label{Tab:N}
\end{table}

\par 
In Table~\ref{Tab:A} the point of maximum~$\tau_A$ and the corresponding value~$A_{max}$ of the curvature variable~$A$ at this point is given as a function of the initial value~$K_0$ of the diffusion term for Bianchi cosmologies of  types~III and~V filled with dust and radiation. For example, for a radiation-filled Bianchi type~III cosmology without diffusion~$A(\tau)$ reaches its maximal value~$A_{max}=0.137$ at~$\tau_A=0.66$, while under conditions of relatively strong diffusion~(${K_0=0.08}$) the maximal value is around~$33\%$ higher~($A_{max}=0.182$) and is reached significantly later~(${\tau_A=1.48}$). Observed are the two general tendencies. Firstly, the diffusion effects are significantly stronger in the models filled with radiation compared to those with dust. Also, more complicated type~III and~IV universes are more subject to diffusional effects than less advanced type~V cosmologies.
\par 
Table~\ref{Tab:N} shows the corresponding results for the curvature variable~$N$. Similarly, the maximal value~$N_{max}$ increases and is reached at a later time moment as the diffusion intensity grows.

\begin{table}[ht]
\centering
\begin{tabular}{|c|c|c|c|c|c|c|c|c|c|c|}
\hline 
Quantity			& \multicolumn{4}{c|}{$\Sigma_+(K_0)/\Sigma_+(0)$} & \multicolumn{4}{c|}{$A(K_0)/A(0)$} \\
\hline	
Cosmology		&	\multicolumn{2}{c|}{Type III} 	&		\multicolumn{2}{c|}{Type V} & \multicolumn{2}{c|}{Type III} & \multicolumn{2}{c|}{Type V} \\
\hline 
$\gamma$			&	1			&	4/3		&	1			&	4/3		&	1			&	4/3		&	1			&	4/3\\
\hline 
$K_0=0.02$		&	1.2		&	1.2		&	1.1		&	1.1		&	1.1		&	1.1		&	1.1		&	1.1	\\
\hline 
$K_0=0.04$		&	1.6		&	1.7		&	1.2		&	1.3		&	1.3		&	1.3		&	1.2		&	1.3	\\
\hline
$K_0=0.06$		&	2.2		&	2.7		&	1.4		&	1.5		&	1.5		&	1.7		&	1.4		&	1.5	\\
\hline 
$K_0=0.08$		&	4.4		&	27.8		&	1.7		&	2.1		&	2.2		&	6.4		&	1.7		&	2.1	\\
\hline
\end{tabular}
\caption{Future asymptotic values of ratios~$\Sigma_+(K_0)/\Sigma_+(0)$ and~$A(K_0)/A(0)$ for Bianchi type~III and type~V models filled with dust and radiation}
\label{Tab:PropConstants}
\end{table}

\par 
There can be other ways to describe the diffusional effects. At a fixed timepoint the geometrical variables achieve greater absolute values at stronger diffusion, their future asymptotic form remaining unchanged. So, diffusion affects the proportionality constants such as
\begin{equation}
\frac{Y_G (K_{02})}{Y_G(K_{01})}\to const >1
\end{equation}
at sufficiently late times, if~$K_{02}>K_{01}.$ The future asymptotic values of the ratios~$\Sigma_+(K_0)/\Sigma_+(0)$ and~$A(K_0)/A(0)$ for dust- and radiation-filled Bianchi type~III and type~V universes are given in Table~\ref{Tab:PropConstants}. It can be found, for example, that in a dust-filled type~III universe at a fixed moment of time in a far enough future the shear~$\Sigma_+$ is~$4.4$~times higher, and the curvature variable~$A$ is~$2.2$~times higher than in the same model without diffusion. A significantly greater value of~$\Sigma_+(0.08)/\Sigma_+(0)$ for Bianchi type~III,~$\gamma=4/3$ is explained by the vicinity of~$K_0=0.08$ to the critical value~$K_r$ for this model.

\subsection{Past dynamics of the models}
Qualitatively, presence of diffusion affects neither the past asymptotic state of the considered models nor the past dynamics of the universe. Independently of the fluid type, at earlier times the solution of the system is asymptotically an extremely tilted Kasner solution:
\begin{align}
\begin{split}
(\Sigma_+,\Sigma_-,\Sigma_{12},\Sigma_{13},\Sigma_{23},A,N) &\to (\Sigma_+^*,\Sigma_-^*,0,0,0,0,0), \\
(\Omega,\Phi,K,v_1,v_2,v_3)& \to (0,0,0,1,0,0), \\
{(\Sigma_+^*)}^2+{(\Sigma_-^*)}^2 &= 1.
\end{split}
\end{align}

Quantitatively, the effect of diffusion on the past dynamics is very subtle. For example, the values of~$\Omega$ and~$\Phi$ at~$-1<\tau<0$ in the case with diffusion are slightly different from those in the diffusionless case. This can be seen in Figures~\ref{Fig:Omegas} and~\ref{Fig:Phis}.

\subsection{The asymptotic value of~$\lambda$ in Bianchi type III models}
\label{Sec:Results:Lambda}
As the geometry of the model evolves towards de Sitter, the~$\lambda$-variable tends to some positive constant value:~$\lambda \to \lambda_\infty,~0<\lambda_\infty<1$. The asymptotic value of~$\lambda$ is affected by the initial value of diffusion term; namely,~$\lambda_\infty$ is reduced as~$K_0$ grows.

\par 
The behaviour of~$\lambda$ at positive times for dust- and radiation-filled Bianchi type~III universes is demonstrated in Figure~\ref{Fig:Lambda}. The solid lines correspond to initial values of the diffusion term~$K_0=0.02; \, 0.04; \, 0.06; \, 0.08$ (the higher~$K_0$, the stronger the reduction), while the dynamics in the diffusionless case is shown by the dashed line. It can be seen that the impact of diffusion on radiation-filled models is greater and leads to a more significant reduction in~$\lambda_\infty$.

\begin{figure}[ht]
\begin{minipage}[h]{0.49\linewidth}
\includegraphics[width=1.0\linewidth]{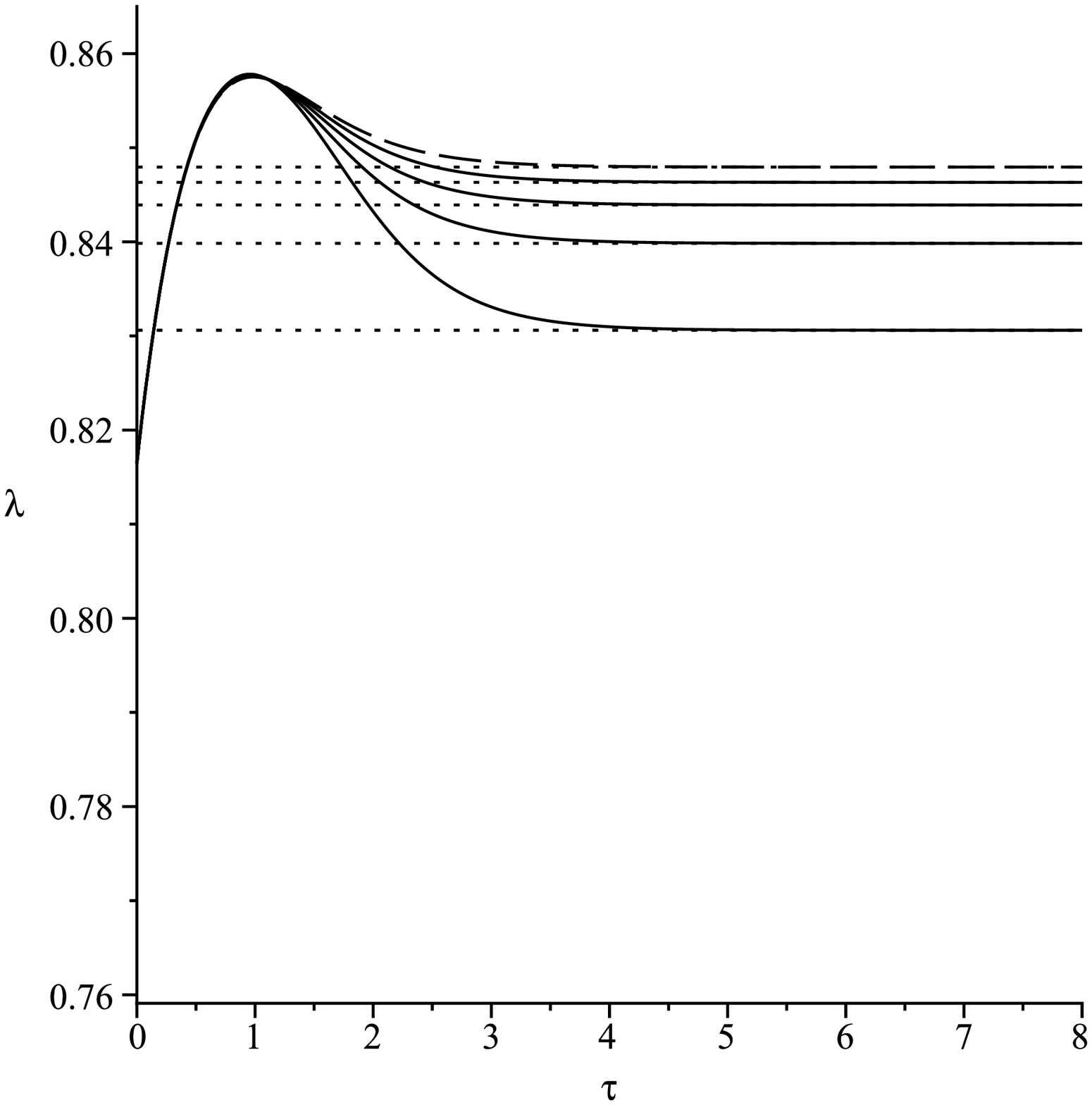}
\end{minipage}
\begin{minipage}[h]{0.49\linewidth}
\includegraphics[width=1.0\linewidth]{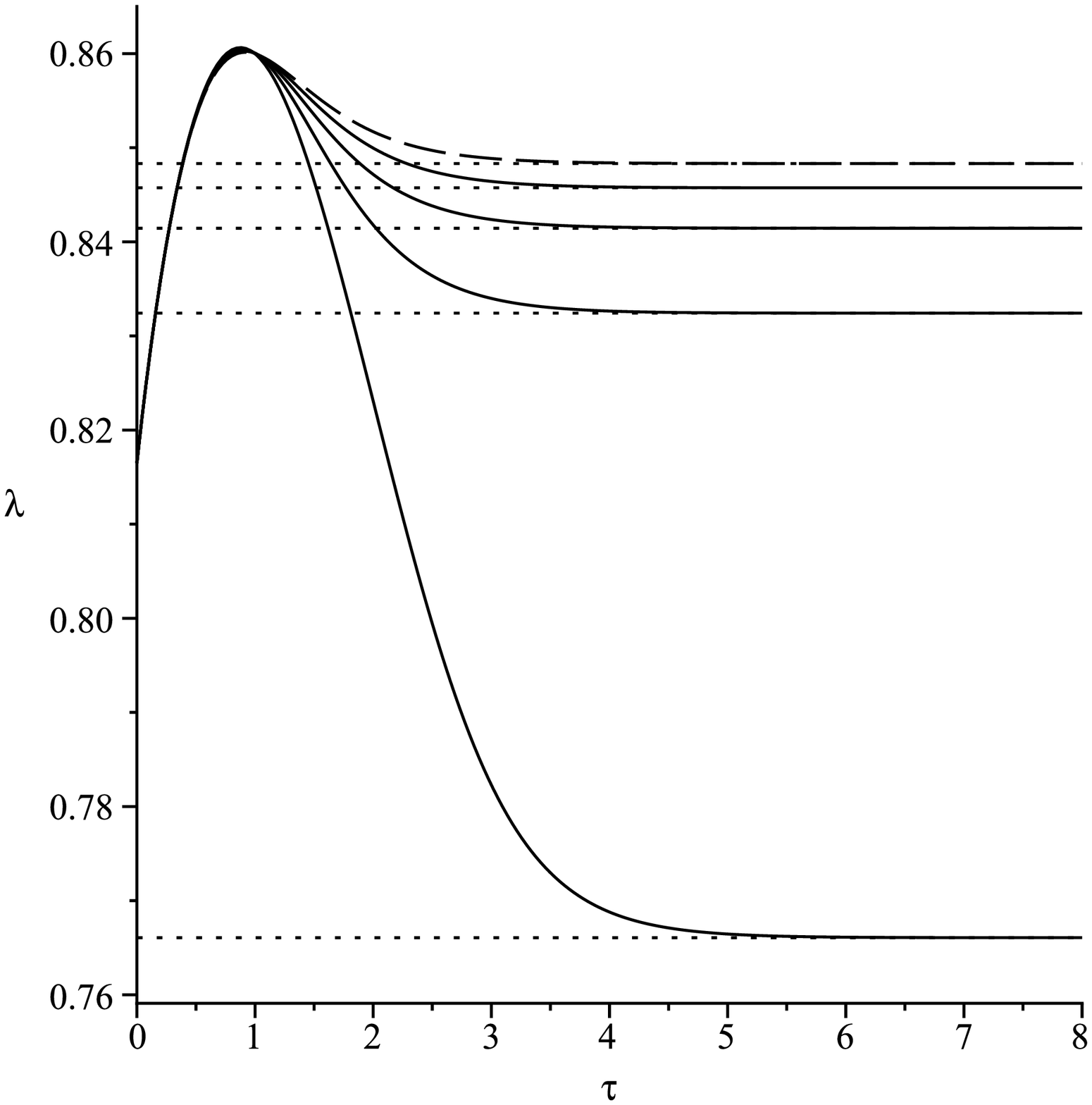}
\end{minipage}
\caption{Dynamics of~$\lambda$ for Bianchi type III models filled with dust (left) and radiation (right) for~$K_0=0; \, 0.02; \, 0.04; \, 0.06; \, 0.08$}
\label{Fig:Lambda}
\end{figure}

\par 
The numerical data are demonstrated in Table~\ref{Tab:Lambda}. The table shows how the future asymptotic value of~$\lambda$ varies as a function of~$K_0$ for type~III cosmologies filled with dust and radiation. Namely, at small initial values of the diffusion term (for example,~$K_0=0.02$) the change in~$\lambda_\infty$ is insignificant and does not exceed 0.5\%, while the reduction becomes more substantial as~$K_0$ grows and approaches~$K_r$. For example, for a dust-filled universe~${\lambda_\infty=0.7044}$ at~$K_0\to K_r$, while~$\lambda_\infty=0.8479$ in the absence of diffusion. That is, strong diffusion reduces the ultimate value of~$\lambda$ approximately by~17\%. The reduction is even larger for a radiation-filled universe (around 22\%).

\begin{table}[ht]
\centering
\begin{tabular}{|c|c|c|c|c|c|c|c|c|c|c|}
\hline 
$K_0$ 	 	 & 0 & 0.02 & 0.04 & 0.06 & 0.08 & 0.09 & $K_r$ \\
\hline 
$\gamma=1$	 & 0.8479 & 0.8463 &  0.8439 & 0.8398 & 0.8306 & 0.8164	& 0.7044 \\
\hline 
$\gamma=4/3$ & 0.8483 & 0.8457 &  0.8414 & 0.8324 & 0.7661 & N/A		& 0.6644 \\
\hline 
\end{tabular}
\caption{Future asymptotic values of~$\lambda$ in Bianchi type~III universes for the two kinds of fluid (dust and radation) as functions of the initial value~$K_0$ of the diffusion term. The last column shows the case~$K_0\to K_r$, which we assume to be achived at~$K_r-K_0<10^{-4}$ from the numerical point of view}
\label{Tab:Lambda}
\end{table}

\subsection{Special dynamics of shear variables in type~III and type~IV models}
\label{Sec:Results:Deformations}

The simulation has shown that diffusion has a special impact on dynamics of the shear variables in Bianchi cosmologies of types~III and~IV. For example, while in the diffusionless case~$\Sigma_+(\tau)$ and~$\Sigma_-(\tau)$ are monotone functions and in addition~$\Sigma_-(\tau)$ has no inflection points, the change of sign of the derivatives~$\Sigma_+^\prime(\tau),~\Sigma_-^{\prime\prime}(\tau)$ is observed already at moderate values of~$K_0$. This may be explained by the increase in the curvature variables caused by diffusion (see Section~\ref{Sec:Results:Quantitative}). Without diffusion similar behaviour can be found only at substantial deviations from the Robertson-Walker-approaching initial conditions~(\ref{Eq:IC:Inconds}). This effect is present both for dust- and radiation-filled cosmologies, but appears at lesser values of~$K_0$ at~$\gamma=4/3$.
\par 
The described deformations are absent for more simple type~V cosmologies. Both at extremely intense diffusion~$(K_0\to K_r)$ and at alteration of initial conditions the dynamics of the shear variables at positive times is described by monotone functions without inflection points.

\begin{figure}[ht]
\begin{minipage}[h]{0.49\linewidth}
\includegraphics[width=1.0\linewidth]{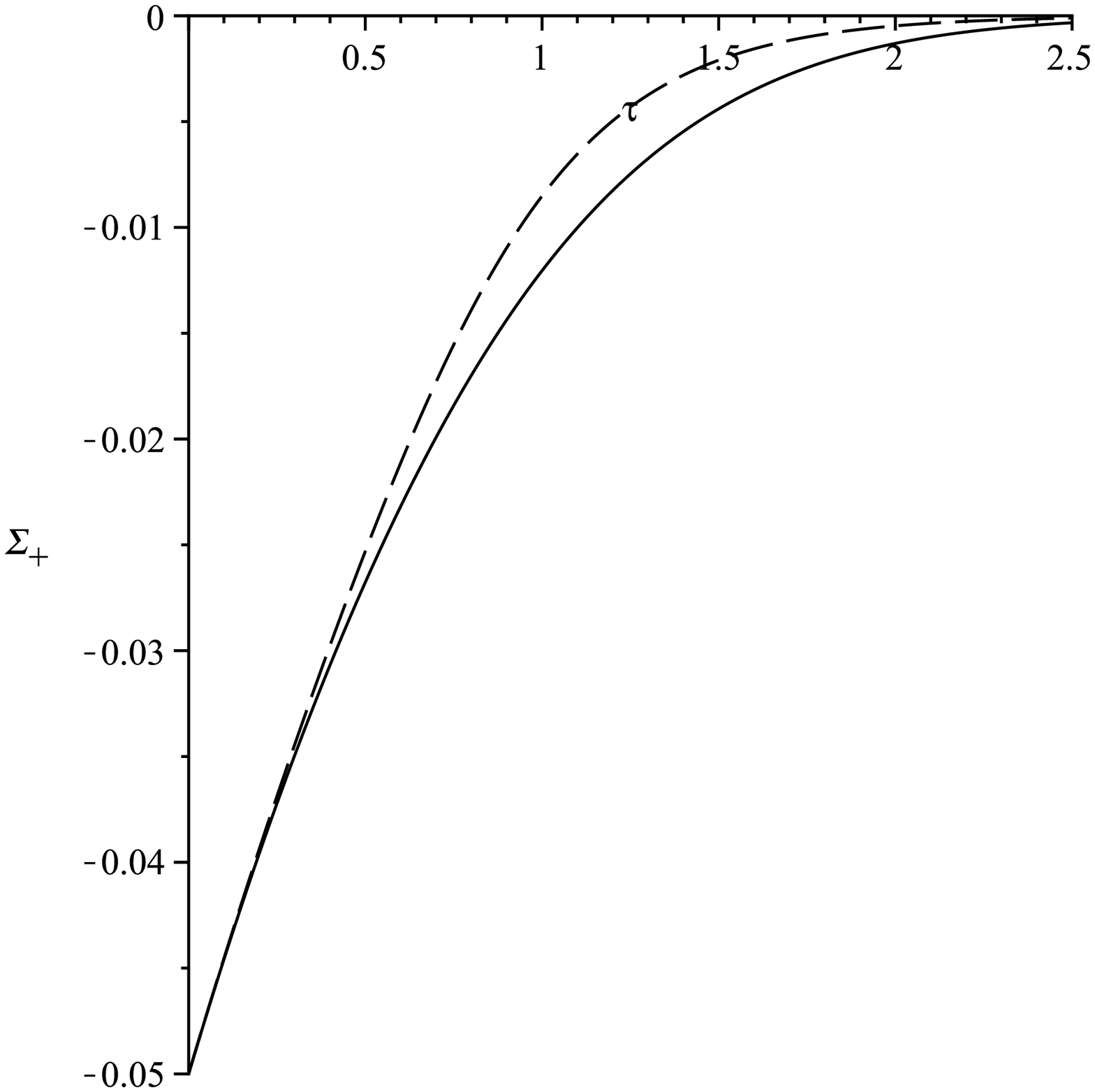}
\end{minipage}
\begin{minipage}[h]{0.49\linewidth}
\includegraphics[width=1.0\linewidth]{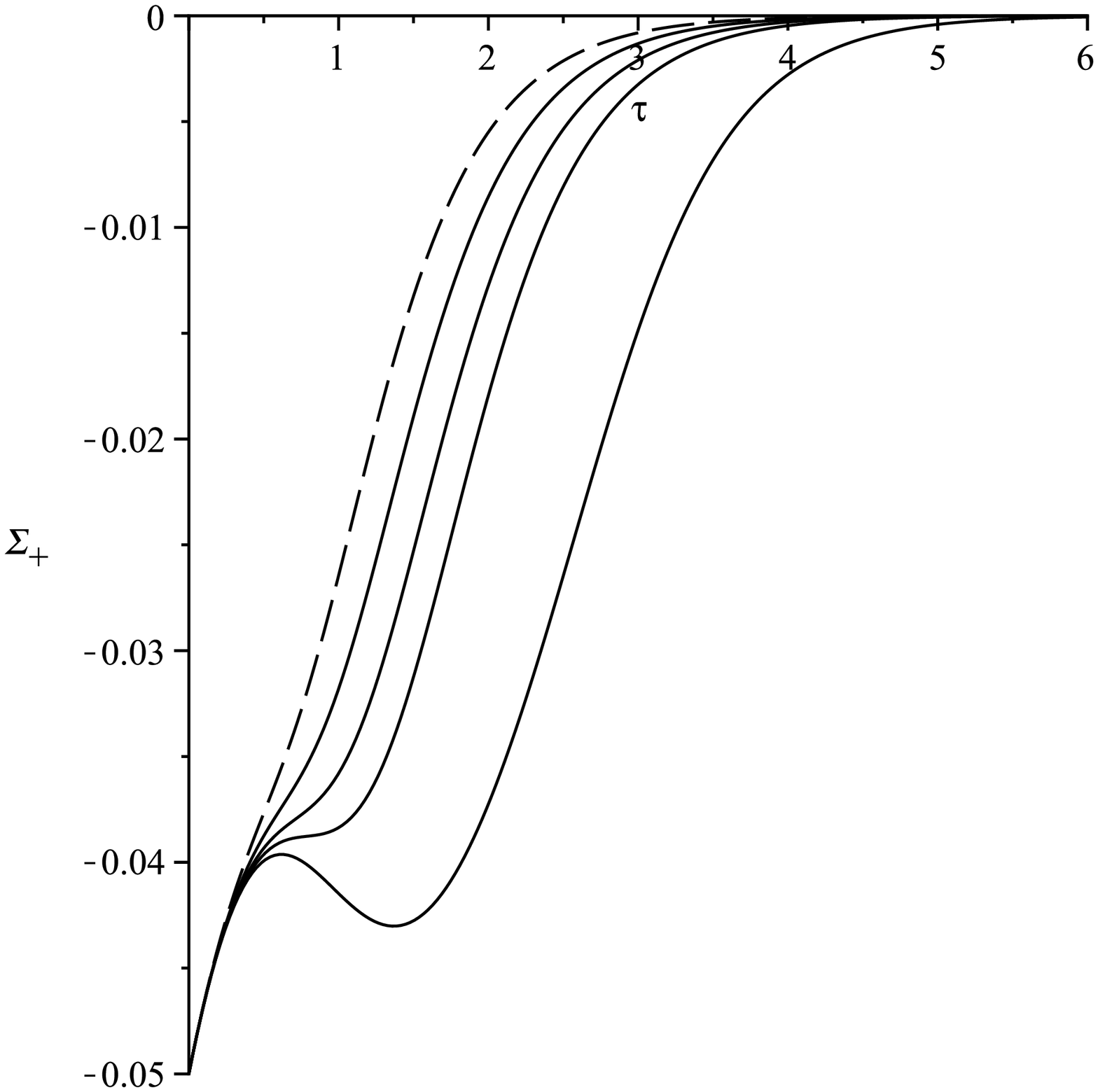}
\end{minipage}
\caption{Dynamics of the shear variable~$\Sigma_+$ for radiation-filled models: Bianchi type~V at~$K_0=0$ and~$K_0 \to K_r$~(left) and type~III at~${K_0=0; \, 0.02; \, 0.04; \, 0.06; \, 0.08}$~(right)}
\label{Fig:SigPlus}
\end{figure}

\par 
In Figure~\ref{Fig:SigPlus} one can see how differently the increase of diffusion term changes the dynamics of the shear variable~$\Sigma_+$ in various models. In type~V universes (left) the only difference between the diffusionless case (the dashed line) and the case of extreme diffusion~$K_0\to K_r$ (the solid line) is only quantitative. Another situation is observed for type~III models (right). The solid lines here correspond to~${K_0=0.02;\,0.04;\,0.06;\,0.08}$ from left to right. It can be seen that the inflections become more distinct as diffusion increases, and at~$0.06<K_0<0.08$ the monotone behaviour of~$\Sigma_+(\tau)$ is broken. A maximum and a minimum of the function are clearly observed at~$K_0=0.08$.

\begin{figure}[ht]
\begin{minipage}[h]{0.49\linewidth}
\includegraphics[width=1.0\linewidth]{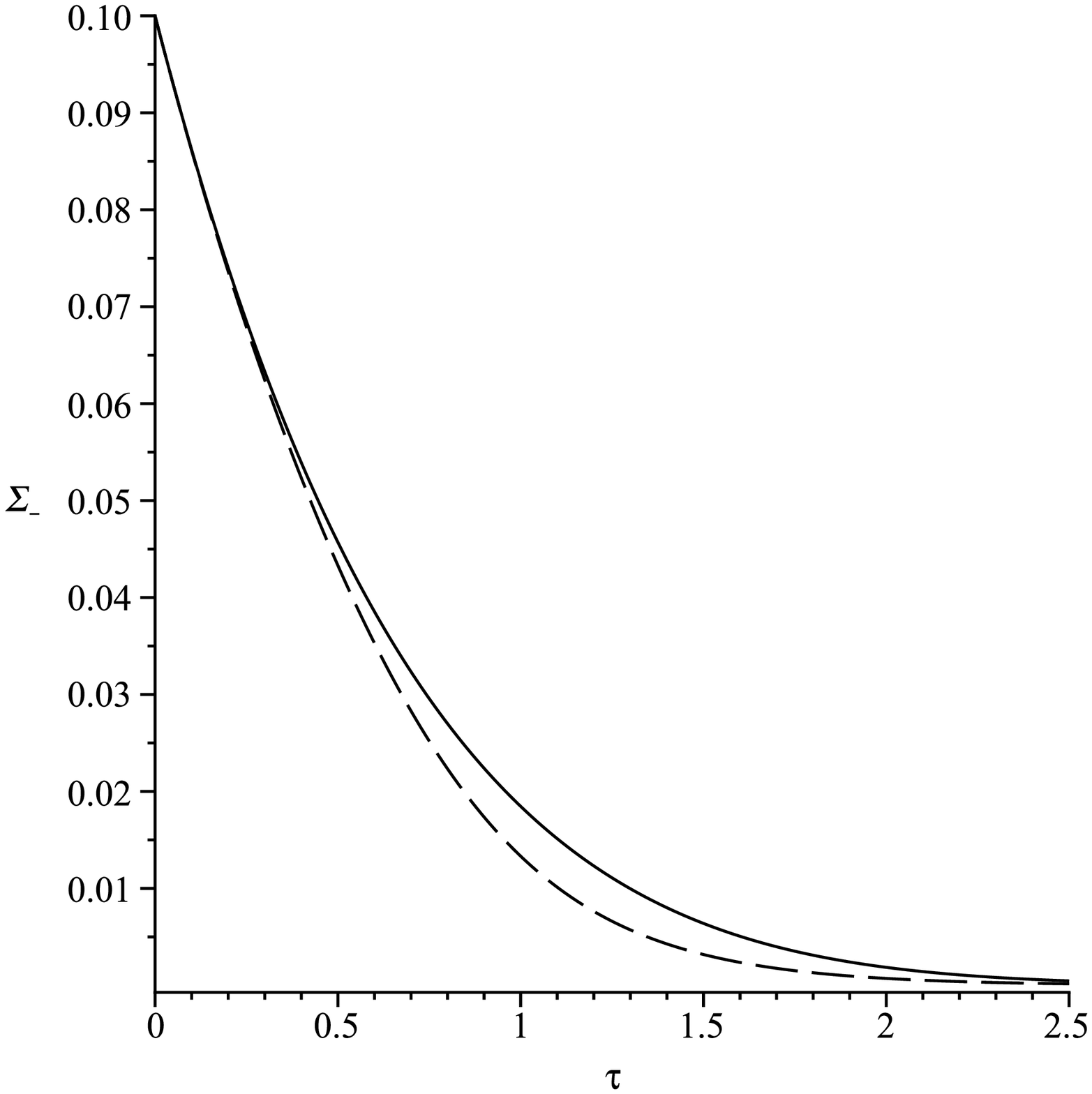}
\end{minipage}
\begin{minipage}[h]{0.49\linewidth}
\includegraphics[width=1.0\linewidth]{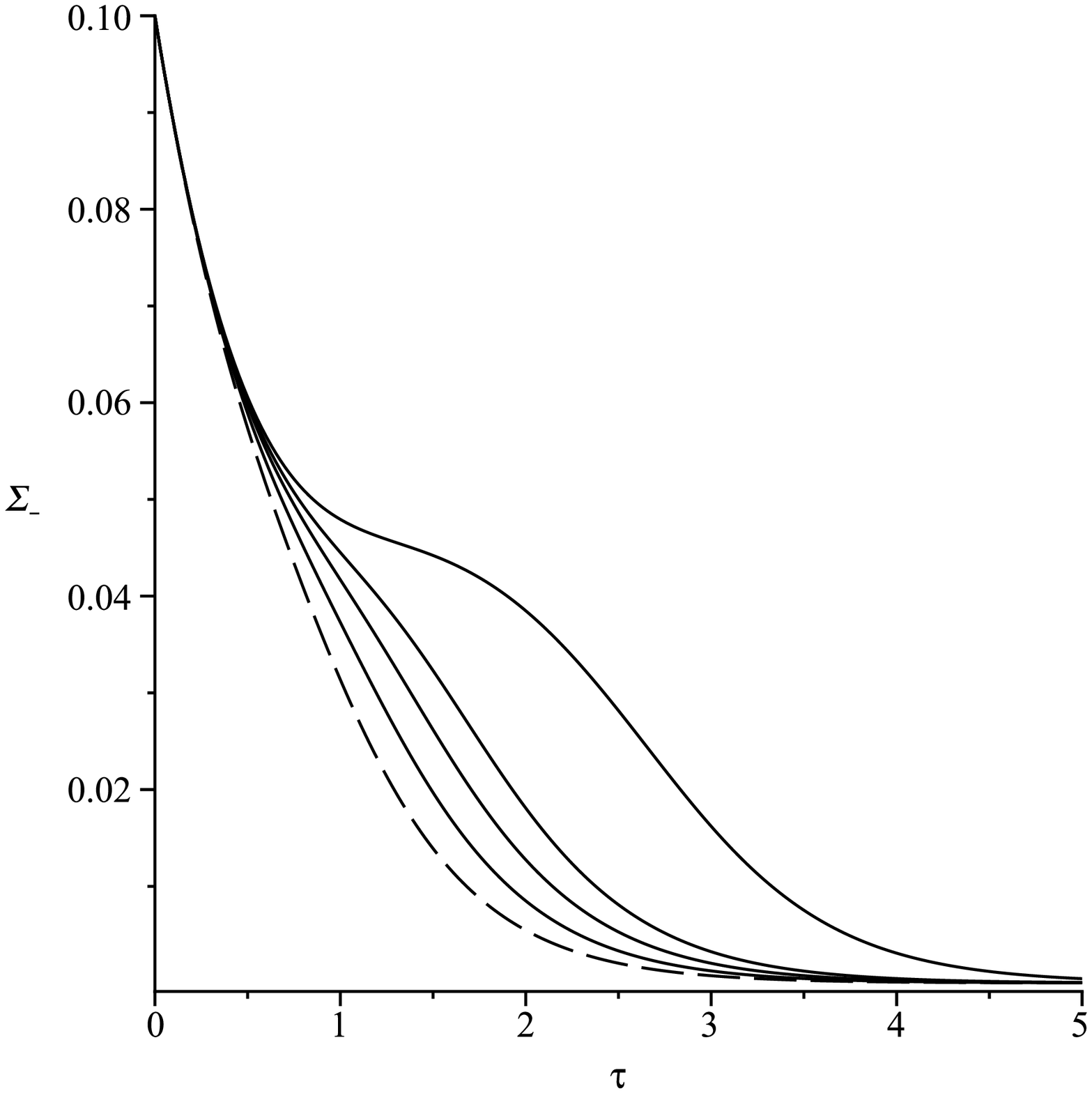}
\end{minipage}
\caption{Dynamics of the shear variable~$\Sigma_-$ for different cosmologies at~$\gamma=4/3$: Bianchi type~V at~$K_0=0$ and~$K_0 \to K_r$~(left) and Bianchi type~III at~${K_0=0; \, 0.02; \, 0.04; \, 0.06; \, 0.08}$~(right)}
\label{Fig:SigMinus}
\end{figure}

\par 
Similarly, the dynamics of~$\Sigma_-$ for radiation-filled cosmologies is shown in Figure~\ref{Fig:SigMinus}. Again, the diffusion has only quantitative impact on Bianchi type~V (left), where the dashed line corresponds to the case without diffusion and the solid line to~$K_0\to K_r.$ For Bianchi type~III models (right) the curves show the cases~$K_0=0; \, 0.02; \, 0.04; \, 0.06; \, 0.08$ (from left to right, starting from the dashed line for~$K_0=0$). It can be seen that inflections are initially absent but appear already at~$K_0=0.04$ and become more clear at greater values of the diffusion term. 

\section{Summary}
\label{Sec:Summary}
We have used the dynamical systems approach to perform an analysis of the dynamics of fully tilted dust- and radiation-filled Bianchi models of types~III,~IV, and~V in presence of diffusion. In particular, type~V cosmologies appear to be less subject to the diffusional effects than the other two types under consideration. 
\par 
Aside from the typical quantitative differences which we have described in details, diffusion can drastically change the dynamics of Bianchi cosmologies from a qualitative point of view. The cosmic no-hair theorem is not valid for the models with diffusion, and future recollapse of the universe becomes possible if diffusion is strong enough, in contrast to the standard diffusion-free models with a positive cosmological constant. When recollapse does not happen, the models evolve towards the de Sitter state of accelerated expansion.
\par 
Another key result is the isotropization of radiation in Bianchi models with diffusion. We have shown that at late times radiation becomes asymptotically non-tilted, while this never happens in the diffusionless case; namely, the tilt components decrease exponentially at later times with~${v_1,~v_2,~v_3\sim e^{-0.25\tau}}$. At the same time, the tilt dynamics for dust is not significantly affected by diffusion.
\par 
Also, we have calculated the decay rates of the energy density in models with diffusion and shown that this quantity decreases slower compared to the case when diffusion is absent:
\begin{equation}
\left\{ \begin{array}{lll} 
\Omega_{rad} \sim e^{-4\tau}, & \Omega_{dust}\sim e^{-3\tau} & \text{without diffusion}; \\
\Omega_{rad} \sim e^{-3\tau}, & \Omega_{dust}\to (C_K\cdot \tau+C_\Omega)e^{-3\tau}, & \text{with diffusion}.
\end{array} \right.
\end{equation}
On the contrary, the future asymptotic behaviour of the scalar potential and the geometrical variables is not significantly changed in presence of diffusion.
\par 
The diffusion term asymptotically approaches zero both in the past and in the future, but experiences a maximum at some positive timepoint. At later times the diffusion term has been found to experience an exponential decay,~$K\sim e^{-3\tau}$.
\par 
In type~III cosmologies, diffusion has been shown to affect the future asymptotic value of the $\lambda$-variable. Namely, the stronger the diffusion, the more significantly~$\lambda_\infty$ is reduced.
\par 
At relatively high initial values of diffusion some changes appear in dynamics of the shear variables in cosmologies of types~III and~IV. The found behaviour is observed in the absence of diffusion only at substantial deviations from the Robertson-Walker-approaching initial conditions. At the same time, this effect has no analogue in Bianchi type~V models.   

\begin{acknowledgments}
We would like to thank the reviewers of Classical and Quantum Gravity, whose useful comments helped us to improve the quality of the present manuscript.
\end{acknowledgments}

\bibliography{Bib/New}
\end{document}